\documentclass[11pt]{article} 

\usepackage{amsmath,epsfig}
\usepackage{latexsym}
\usepackage{amsfonts}
\usepackage{graphicx} 
\usepackage{subfigure}
\usepackage{amssymb}
\usepackage{stmaryrd}
\usepackage{caption}
\usepackage{color}
\usepackage{cite,multicol,tabularx}

\newcommand{\dfn}{\triangleq} 

\newcommand{\rw}{\rightarrow}

\newcommand{\mX}{{\mathcal X}}

\newcommand{\mU}{{\mathcal U}}
\newcommand{\mK}{{\mathcal K}}
\newcommand{\mT}{{\mathcal T}}
\newcommand{\Real}{\mathbb{R}}
\newcommand{\sS}{{\sf S}}

\newcommand{\sfh}{{\sf h}}
\newcommand{\sfm}{{\sf m}}
\newcommand{\sd}{{\sf d}}
\newcommand{\sw}{{\sf w}}

\newcommand{\mbE}{\mathbb{E}}
\newcommand{\mbI}{\mathbb{I}}

\newcommand{\bx}{{\bf x}}
\newcommand{\by}{{\bf y}}

\newcommand{\qed}{$\blacksquare$}

\newcommand{\bftheta}{{\mbox{\boldmath $\theta$}}}
\newcommand{\bfepsilon}{{\mbox{\boldmath $\epsilon$}}}

\newtheorem{teorema}{\bf Theorem}
\newtheorem{lema}{\bf Lemma}

\newtheorem{hipotesis}{\bf Assumption}
\newtheorem{nota}{\bf Remark}

\title{An iterative importance sampler for Bayesian parameter estimation in stochastic models of multicellular clocks}

\author{In\'es P. Mari\~no$^1$, Joaqu\'{\i}n M\'{\i}guez$^2$ and Alexey Zaikin$^3$\\
$^1$ Department of Biology and Geology, Physics and Inorganic Chemistry, \\Universidad Rey Juan Carlos.\\
 Tulip\'an s/n, 28933 M\'ostoles, Madrid (Spain). Email: {\sf ines.perez@urjc.es}\\
$^2$ School of Mathematical Sciences, Queen Mary University of London. \\
Mile End Rd, E1 4NS London (UK). E-mail: {\tt j.miguez@qmul.ac.uk}.\\
$^3$ Institute for Women's Health and Department of Mathematics,\\ 
University College London. \\ London W1T 7DN, United Kingdom. E-mail: {\tt alexey.zaikin@ucl.ac.uk}
}
\date{\today}
\begin{document}

\maketitle

\abstract{
We investigate a stochastic version of the synthetic multicellular clock model proposed by Garcia-Ojalvo, Elowitz and Strogatz. By introducing dynamical noise in the model and assuming that the partial observations of the system can be contaminated by additive noise, we enable a principled mechanism to represent experimental uncertainties in the synthesis of the multicellular system and  pave the way for the design of probabilistic methods for the estimation of any unknowns in the model. Within this setup, we investigate the use of an iterative importance sampling scheme, termed nonlinear population Monte Carlo (NPMC), for the Bayesian estimation of the model parameters. The algorithm yields a stochastic approximation of the posterior probability distribution of the unknown parameters given the available data (partial and possibly noisy observations). We prove a new theoretical result for this algorithm, which indicates that the approximations converge almost surely to the actual distributions, even when the weights in the importance sampling scheme cannot be computed exactly. We also provide a detailed numerical assessment of the stochastic multicellular model and the accuracy of the proposed NPMC algorithm, including a comparison with the popular particle Metropolis-Hastings algorithm of Andrieu {\em et al.}, 2010, applied to the same model and an approximate Bayesian computation sequential Monte Carlo method introduced by Mari\~no {\em et al.}, 2013.
}

\vspace{+0.5cm}

\noindent {\bf Keywords:} repressilator; intercellular networks; population Monte Carlo; ...

\addtolength{\baselineskip}{0.25cm}

%
\section{Introduction}

The field of systems biology is rich in problems that demand sophisticated computational tools for estimation, detection and prediction. With the genomics revolution and rise of systems biology, we are witnessing the development of a rigorous engineering discipline to create, control and programme cellular behaviour \cite{Cameron14}. The resulting field, known as synthetic biology, has undergone a dramatic growth throughout the past decade and is poised to transform biotechnology and medicine.  A core issue in synthetic biology is the analysis of
networks of interacting biomolecules \cite{Chabrier04},  which  carry out many essential functions in living cells. However,  the design principles underlying the functioning of such intracellular networks remain poorly understood. It is expected, though, that the ability to design synthetic networks will lead both to the engineering of new cellular behaviours and to an improved understanding of naturally occurring networks. A particular system that has drawn considerable attention is the so-called repressilator \cite{Elowitz00} which is an oscillating network that periodically induces the synthesis of a green fluorescent protein as a readout of its state in individual cells and can be considered as synthetic biological clock. Mathematical models, consisting of systems of nonlinear differential equations, that describe the dynamics of the original repressilator and subsequent extensions of it have appeared in the literature \cite{Elowitz00,Garcia-Ojalvo04,Ullner07,Ullner08,Koseska10,Marinho13,Evans14} and sparked interest from research in physics, engineering and mathematics.

 In this paper we investigate the application of a novel method termed nonlinear population Monte Carlo (NPMC) \cite{Koblents14} to the estimation of a subset of the static parameters of a  modified version of this repressilator model \cite{Garcia-Ojalvo04,Ullner07,Marinho13}. 
In particular, we introduce a stochastic version of the chaotic, continuous-time modified repressilator model of \cite{Marinho13}, which consists of a set of stochastic differential equations (SDEs) driven by Wiener noise process. These equations depend on a number of unknown parameters, which we model as random variables. We convert the system of SDEs into an (approximate) discrete-time state space model using a standard Euler-Maruyama scheme and then consider the problem of computing the posterior probability distribution of the unknown parameters in the model conditional on a sequence of partial observations that consist of noisy measurements of a small subset of the (dynamic) state variables. This setup resembles the scenario considered in \cite{Marinho13} but (i) the system in this paper is stochastic, while in \cite{Marinho13} only a deterministic model was studied, and (ii) we pose a data-poor problem, with the collected observations being low dimensional (2-dimensional, for a 14-dimensional state space), noisy and sparse in time, whereas in \cite{Marinho13} data were assumed available continuously in time and noise-free. The randomness in the proposed model dynamics can potentially account for experimental uncertainties in the synthesis of the biological system.

We tackle the Bayesian estimation of the unknown model parameters, i.e., the approximation of their posterior probability distribution conditional on the available observations. We propose to apply the recently introduced NPMC methodology of \cite{Koblents14}, which involves an iterative importance sampling scheme where the weights undergo a nonlinear transformation to control their variance and mitigate the well-known degeneracy problem of importance samplers \cite{Doucet00,Koblents14}. In \cite{Koblents14} it was proved that the estimates produced at each iteration of an  NPMC algorithm converge in probability 
and almost surely (a.s.) as the number of Monte Carlo samples, $M$, increases.
Therefore, the weight transformation improves the finite-sample-size performance of the importance sampling scheme, while preserving asymptotic convergence.
 In the problem tackled in the present paper, however, the weights cannot be computed exactly, but only approximated using a standard bootstrap filter (see, e.g., \cite{Doucet00}), in a way similar to the particle Markov chain Monte Carlo (PMCMC) method \cite{Andrieu10} and some of the numerical examples in \cite{Koblents14}. Based on certain unbiasedness properties of particle filters we rigorously prove that the importance sampler with transformed weights
 also attains asymptotic convergence when using {\it approximate} weights, even if the complexity of the bootstrap filters used for the approximations is fixed (and hence they introduce errors that do not vanish asymptotically). The latter feature is known as {\em exact approximation} in the PMCMC literature. Unlike the theorem in \cite{Koblents14}, the analysis in this paper yields an explicit almost sure convergence rate of the form $M^{-\frac{1}{2}+\epsilon}$, where $M$ is the number of MC samples and $\epsilon>0$ is an arbitrarily small constant.
 We have run computer simulations to illustrate how the parameters of the stochastic modified repressilator can be estimated using the proposed NPMC method with noisy partial observations. We have also compared the relative performance of the NPMC method, the approximate Bayesian computation  sequential Monte Carlo (ABC SMC) technique  \cite{Marinho13,Toni09} and a particle Metropolis-Hastings (PMH) algorithm. Our results show that the NPMC method is more accurate than the state of the art PMH algorithm or the ABC SMC technique with a similar, or even lighter, computational cost.

The rest of the paper is organised as follows. In Section \ref{sRepressilator} we introduce a stochastic coupled-repressilator model of intercellular networks and derive a discrete-time version. The NPMC algorithm for this system is outlined in Section \ref{sNPMC}, and a novel theoretical argument for asymptotic convergence is  introduced in Section \ref{sConvergence}. Some computer simulation results are shown in Section \ref{sSimulations} and, finally, Section \ref{sConclusions} is devoted to the conclusions.

%
\section{Intercellular network model} \label{sRepressilator}
\subsection{Modified stochastic repressilator} \label{stoch}

\begin{figure*}[htb]
\centerline{
	\includegraphics[width=0.5\linewidth,angle=90]{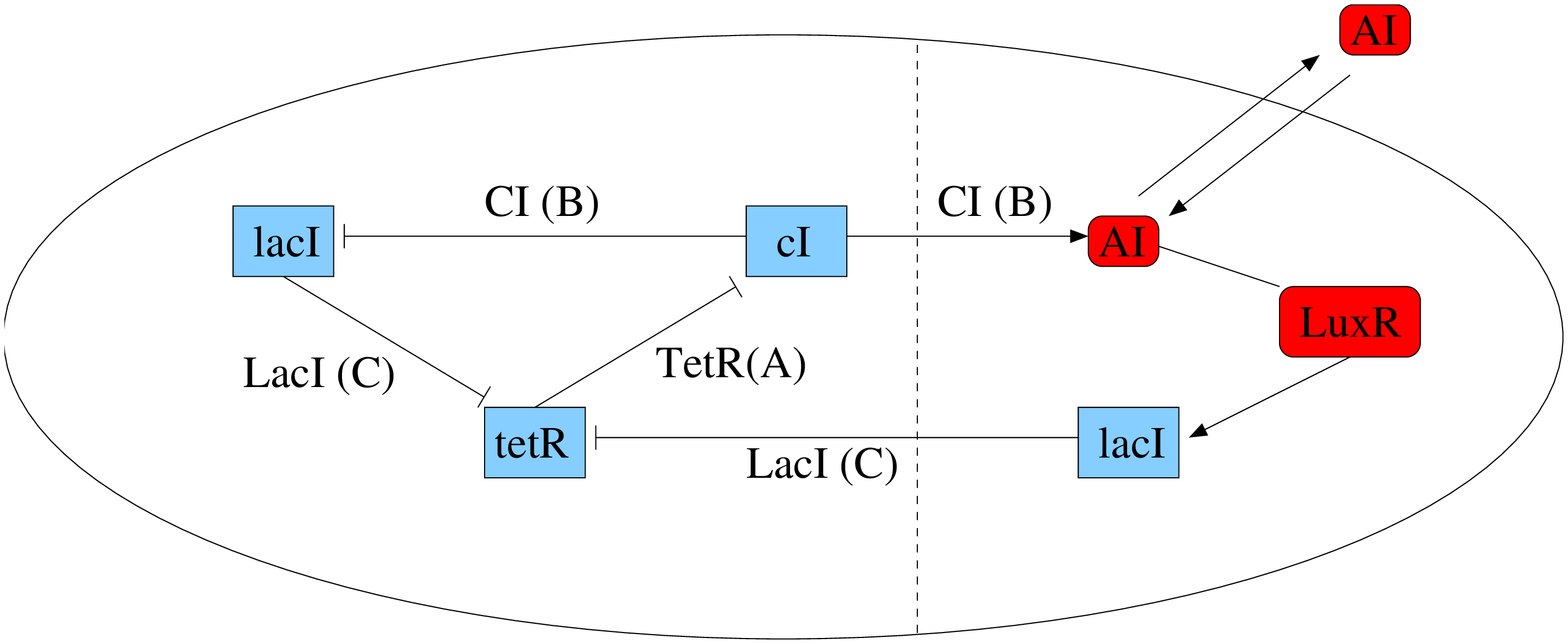}
	}
\caption{Modified repressilator, where the modification is shown in the right side of the plot. Genes and molecules are represented in light blue and red colours, respectively. The oval depicts the cell membrane.}
\label{repressilator_fig}
\end{figure*}

The standard repressilator is a ``genetic clock'' built by three genes, where the protein product of each gene represses the expression of another one in a cyclic manner \cite{Ullner07}. It produces nearly harmonic oscillations in protein levels. In the original repressilator design, the gene {\it lacI} expresses protein {\sf LacI}, which inhibits transcription of the gene {\it tetR}. The product of the latter, {\sf TetR}, inhibits transcription of the gene {\it cI}. Finally, the protein product {\sf CI} of the gene {\it cI} inhibits expression of {\it lacI} and completes the cycle. We can see this mechanism in the left side of Figure \ref{repressilator_fig}, where the genes are represented in light blue colour. 

Figure  \ref{repressilator_fig} represents a modification of the  repressilator, introduced in  \cite{Garcia-Ojalvo04}, that includes an additional feedback loop involving a small autoinducer (AI) molecule produced by  {\sf CI} that can diffuse through the cell membrane, and the protein {\sf LuxR}, which responds to the AI by activating the transcription of a second copy of the repressilator gene  {\it lacI}. Placing the gene {\it cI} under inhibitory control of the repressilator protein {\sf TetR} leads to a repressive and phase-repulsive coupling that, in turn, generates rich dynamical patterns, including chaotic oscillations  \cite{Ullner07}. Phase repulsive coupling is common in many biological systems, e.g., in neural activity, in the brain of songbirds or in the respiratory system.

In this paper we study a model consisting of two modified repressilators with identical parameters, driven by Wiener-type noise and coupled by the fast diffusion of the AI across the cell membranes.  The resulting mRNA dynamics in continuous time $t \in \Real$ is described by a stochastic Hill-type equation with coefficient $\sfm$, namely
 \begin{eqnarray}
\sd a_i &=& -\left( 
	a_i - \frac{\alpha}{1+C_i^\sfm} 
\right)\sd t + \sigma_a a_i \sd W^a_i, \label{1eq} \\
\sd b_i &=& -\left(
	b_i - \frac{\alpha}{1+A_i^\sfm}
\right)\sd t + \sigma_b b_i \sd W^b_i, \label{2eq} \\
\sd c_i &=& -\left(
	c_i - \frac{\alpha}{1+B_i^\sfm} - \frac{\kappa S_i}{1+S_i}
\right) \sd t + \sigma_c c_i \sd W^c_i , \label{3eq}
\end{eqnarray}
where the subscript $i=1, 2$ specifies the cell; $a_i$, $b_i$, and $c_i$ are time-varying state variables (stochastic processes) representing the concentrations of mRNA molecules transcribed from the genes of {\it tetR}, {\it cI}, and {\it lacI}, respectively; the constant parameter $\alpha$ is the dimensionless transcription rate in the absence of a repressor; the constant parameter $\kappa$ is the maximum transcription rate of the {\sf LuxR} promoter; $S_i$ is a state variable representing the concentration of the AI molecule inside cell $i$, and $W^a_i$, $W^b_i$, $W^c_i$, $i=1,2$, are independent standard Wiener processes scaled by the constant non-negative factors $\sigma_a$, $\sigma_b$ and $\sigma_c$, respectively. The additional time-varying states  $A_i$, $B_i$, and $C_i$, $i=1,2$, are stochastic processes representing the concentration of the proteins {\sf TetR}, {\sf CI}, and {\sf LacI}, respectively, whose dynamics obey the SDEs
\begin{eqnarray}
\sd A_i &=& \beta_a (a_i - A_i) \sd t + \sigma_A A_i \sd W^A_i, \label{4eq}\\
\sd B_i &=& \beta_b (b_i - B_i) \sd t + \sigma_B B_i \sd W^B_i, \label{5eq}\\
\sd C_i &=& \beta_c (c_i - C_i) \sd t + \sigma_C C_i \sd W^C_i. \label{6eq}
\end{eqnarray}
The equations above show that the dynamics of the proteins is linked to the amount of the responsible mRNA, and the constant parameters $\beta_{a}$, $\beta_b$ and $\beta_c$ describe the ratio between mRNA and the protein lifetimes (i.e, the inverse degradation rates). Similar to \eqref{1eq}-\eqref{3eq}, the dynamics is driven by independent standard Wiener processes $W^A_i$, $W^B_i$ and $W^C_i$, $i=1,2$, with constant scale factors $\sigma_A, \sigma_B, \sigma_C \ge 0$. The model is made dimensionless by measuring time in units of the mRNA lifetime (assumed equal for all genes) and the mRNA  and protein levels in units of their Michaelis constant. The mRNA concentrations are additionally rescaled by the ratio of their protein degradation and translation rates \cite{Garcia-Ojalvo04}.

The  term $ \frac{\kappa S_i}{1+S_i}$ on the right-hand side of Eq. \eqref{3eq} represents activated production of {\it lacI} by the AI molecule, whose concentration inside cell $i$ is denoted by $S_i$.  The dynamics of {\sf CI} and {\sf LuxI} can be considered identical, given that their production is controlled by the same protein ({\sf TetR}). Hence, the synthesis of the AI is controlled by the concentration $B_i$ of the protein {\sf CI}.  Taking also into account the intracellular degradation of the AI and its diffusion, the dynamics of $S_i$ is modelled as
\begin{equation}
\sd S_i = -\left( 
	k_{s0} S_i - k_{s1} B_i + \eta(S_i-S_e)
\right) \sd t + \sigma_S S_i \sd W^S_i, 
\label{7eq}
\end{equation}
where $k_{s0}$, $k_{s1}$ and $\eta$ are constant parameters, the latter being a diffusion coefficient that depends on the permeability of the membrane to the AI. The variable $S_e$ is the extracellular concentration of the AI molecule. It is common to apply a quasi-steady-state approximation to the dynamics of $S_e$ \cite{Garcia-Ojalvo04}, which leads to
 $
{S}_e = Q \overline{S} \equiv Q \frac{1}{ N} \sum_{i=1}^N S_i,
$
where $Q = \frac{
	\delta N
}{
	V_{ext} \left(
		k_{se} + \frac{ 
			\delta N
		}{
			V_{ext}
		}
	\right)
}$, $N=2$ is the number of cells, $V _{ext}$ is the total extracellular volume, $k_{se}$ is the extracellular AI degradation rate, and $\delta$ is the product of the membrane permeability and the surface area.

This model can produce a range of dynamic regimes. We achieve an underlying chaotic behaviour for this model when the constant parameters are set as \cite{Ullner07} $Q=0.85$,  $\sfm=2.6$, $\alpha=216$,  $\beta_a=0.85$,  $\beta_b=\beta_c=0.1$, $\eta=2$,  $\kappa=25$, $k_{s0}=1$ and $k_{s1}=0.01$. We will refer to them as standard values.

%
\subsection{Numerical integration and state space model} \label{ssStateSpace}

In order to integrate the 14-dimensional SDE described by Eqs. \eqref{1eq}-\eqref{7eq} numerically, we 
 apply the Euler-Maruyama discretisation with integration step ${\sf h} << 1$, that can be  explicitly written as
 \begin{eqnarray}
a_{i,m+1} &=& a_{i,m}-h\left( 
	a_{i,m} - \frac{\alpha}{1+C_{i,m}^\sfm} 
\right) + \sigma_a a_{i,m}  w^{(1)}_{i,m}, \label{8eq}\\
b_{i,m+1} &=& b_{i,m}-h\left(
	b_i (n)- \frac{\alpha}{1+A_{i,m}^\sfm}
\right) + \sigma_b b_{i,m} w^{(2)}_{i,m}, \label{9eq}\\
 c_{i,m+1} &=& c_{i,m} -h\left(
	c_{i,m} - \frac{\alpha}{1+B_{i,m}^\sfm} - \frac{\kappa S_{i,m}}{1+S_{i,m}}
\right)  + \sigma_c c_{i,m} w^{(3)}_{i,m} ,\label{10eq} \\
 A_{i,m+1} &=& A_{i,m} + h \beta_a (a_{i,m} - A_{i,m})  + \sigma_A A_{i,m} w^{(4)}_{i,m}, \label{11eq}\\
 B_{i,m+1} &=& B_{i,m} + h \beta_b (b_{i,m} - B_{i,m})  + \sigma_B B_{i,m} w^{(5)}_{i,m}, \label{12eq}\\
 C_{i,m+1} &=& C_{i,m} +h \beta_c (c_{i,m} - C_{i,m})  + \sigma_C C_{i,m} w^{(6)}_{i,m}, \label{13eq}\\
 S_{i,m+1} &=& S_{i,m} -h\left( 
	k_{s0} S_{i,m} - k_{s1} B_i(n) + \eta(S_{i,m}-S_{e,m})
\right) + \sigma_S S_{i,m} w^{(7)}_{i,m}\label{14eq}, 
\end{eqnarray}
where $i=1,2$ and $\{w_{i,m}^{(1)}, \ldots, w_{i,m}^{(7)}\}$ are independent Gaussian random variables (r.v.'s) with zero mean and variances $\sigma_a^2, \sigma_b^2, \sigma_c^2, \sigma_A^2, \sigma_B^2, \sigma_C^2$ and $\sigma_S^2$, respectively. 

This system described by Eqs. \eqref{8eq}-\eqref{14eq}  can be compactly written as the multidimensional difference equation
\begin{equation}
\bar\bx_m = F_\theta( \bar\bx_{m-1}, {\bf w}_m ), 
\label{eqDTDE}
\end{equation}
where $F_\theta:\Real^14  \rw \Real^14$ is a function that accounts for both the deterministic and the stochastic part of the model and depends on the parameters $\theta$, $\bar\bx_m=[\bar\bx_{1,m}^\top, \bar\bx_{2,m}^\top]^\top$ is the $14 \times 1$ state of the system at discrete time $m \in \mathbb{Z}$, $\bar\bx_{i,m}$ are the $7 \times 1$ state vectors associated to the two cells, $i=1, 2$, and ${\bf w}_m=[{\bf w}_{1,m}^\top, {\bf w}_{2,m}^\top]^\top$ is an independent and identically distributed (i.i.d.) sequence of $14 \times 1$ zero-mean Gaussian vectors. Each $7 \times 1$ subvector ${\bf w}_{i,m}$, $i=1,2$, has the same distribution and can be written as ${\bf w}_{i,m} = [w_{i,m}^{(1)}, \ldots, w_{i,m}^{(7)}]^\top$.  In particular, note that, for each cell, 
$$
\bar\bx_{i,m} = [a_{i,m},b_{i,m},c_{i,m}, A_{i,m},B_{i,m},C_{i,m},S_{i,m} ]^\top,
$$
with the continuous-time state variables evaluated at time $t=m\sfh$, e.g., $a_{i,m}=a_i(t=m\sfh)$
As in \cite{Marinho13} all constant parameters are assumed known except $\theta=[Q,\sfm,\alpha,\beta_a]^\top$, which are unknown and modelled as random variables.  We assume uniform and independent a priori probability distributions for each one of these parameters, namely $Q\sim \mU(0,1)$, $\sfm \sim \mU(1,5)$, $\alpha \sim \mU(50, 300)$ and $\beta_a \sim \mU(0,1)$. The parameter vector $\theta$, therefore, takes values on the set $\sS = (0,1)\times(1,5)\times(50,300)\times(0,1)$. We denote the conditional (on $\theta$) Markov kernel that determines the state transition from time $m-1$ to time $m$ as $\bar \mK_\theta(\sd \bx_m | \bx_{m-1} )$. In particular, for a Borel set $A \subset \Real^{14}$, $\bar \mK_\theta(A | \bx_{n-1} )$ is the probability of moving from the point $\bar\bx_{m-1}$ in the state space to some  $\bar\bx_m \in A$.

Partial and noisy observations of the system are collected every $m_0$ discrete time steps, i.e., every $t_o=m_0\sfh$ continuous time units. Only the variables $a_i$, $i=1, 2$, are observable, hence the observations have the form
\begin{equation}
\by_n = \left[
	\begin{array}{c}
	a_{1,nm_o}\\
	a_{2,nm_o}\\
	\end{array}
\right] + \sigma_y \bfepsilon_n, \quad n = 1, 2, ...
\label{eqObserv}
\end{equation}
where $\bfepsilon_n$ is a sequence of independent $2 \times 1$ random vectors (with zero mean and identity covariance matrix) and $\sigma_y > 0$ is a known constant parameter.

In order to put the states and the observations on the same time scale, we define the sequence of states $\{ \bx_n \}_{n \ge 0}$ as $\bx_n \dfn \bar\bx_{nm_0}$ and introduce the composite Markov kernel 
\begin{equation}
\mK_\theta(\sd\bx_n|\bx_{n-1}) = \bar\mK_\theta(\sd\bx_n|\bar\bx_{nm_0-1})  
\bar\mK_\theta(\sd\bar\bx_{nm_0-1}|\bar\bx_{nm_0-2}) \cdots 
\bar\mK_\theta(\sd\bar\bx_{(n-1)m_0+1}|\bx_{n-1}).
\label{eqKcomposite}
\end{equation}
For a Borel set $A \subset \Real^{14}$, $\mK_\theta(A|\bx_{n-1})$ is the probability of $\bx_n = \bar\bx_{nm_0} \in A$ conditional on $\bx_{n-1}=\bx_{(n-1)m_0}$ (and the parameter vector $\theta$).

The pair of sequences $\bx_n$ and $\by_n$ yield a discrete-time, Markov state space model \cite{West96} conditional on the choice of parameters $\theta$. The model is specified by the prior probability distribution of the state $\bx_0$, which we denote as $\mK_0(\sd\bx_0)$, the dynamics of the state sequence $\bx_n$, which is given by the Markov kernel $\mK_\theta(\sd\bx_n|\bx_{n-1})$, and the conditional pdf of the observations $\by_n$ given the states $\bx_n$, which we denote as $l_n(\by_n|\bx_n)$. We note that, in this model, the latter density is independent of the parameters $\theta$. Also, since $\by_n-[a_{1,nm_0}, a_{2,nm_0}]^\top = \bfepsilon_n$, the form of the function $l_n(\by_n|\bx_n)$ is given by the pdf of the noise term $\bfepsilon_n$. We often refer to $l_n(\by_n|\bx_n)$ as the likelihood of the state $\bx_n$.

%
\section{Algorithm} \label{sNPMC}

In a Bayesian probabilistic setup, the unknown parameters are modelled as a random vector and the aim is to approximate its posterior probability distribution, conditional on the available observations $\by = \{ \by_1, \by_1, \ldots, \by_R \}$ for some fixed $R>0$. We denote the posterior pdf of the parameters as $p(\theta|\by)$ and note that it can be readily factored, using Bayes' theorem, as $p(\theta|\by) \propto \ell(\by|\theta) p_0(\theta)$, where $\ell(\by|\theta)$ is proportional to the conditional pdf of the observations $\by$ given the parameters $\theta$ (i.e., the likelihood of $\theta$) and $p_0(\theta)$ is the prior density of $\theta$, which has been chosen to be uniform on the set $\sS$, as described in Section \ref{ssStateSpace}. 


The NPMC algorithm of \cite{Koblents14} is an iterative importance sampling (IS) scheme that seeks to approximate a target probability distribution, in our case given by the pdf $p(\theta|\by)$, using weighted Monte Carlo samples. This algorithm generates a sequence of proposal pdf's $q_k(\theta)$, $k = 1, \ldots, K$, from which samples can be drawn and importance weights (IWs) can be computed. This sequence of proposals is expected to yield increasingly better approximations of the target as the algorithm converges. The key feature of the NPMC method, which departs from the classical PMC technique of \cite{Cappe04}, is to compute a set of {\em transformed} importance weights (TIWs) by applying a nonlinear function to the standard IWs. The aim of this transformation is to mitigate the well known problem of the degeneracy of the IWs (common to many IS methods, see \cite{Doucet00,Koblents14}) by controlling the weight variability.

For the case of general state space models, an additional difficulty encountered when trying to estimate the unknown model parameters (denoted $\theta$ in our setup) is that the likelihood $\ell(\by|\theta)$ is intractable. In the last few years, though, it has become a common approach to approximate this likelihood via particle filtering (PF) (see, e.g., \cite{Andrieu10,Chopin12,Koblents14,Maroulas12}). To be specific, we let $\ell^N(\by|\theta)$ stand for the approximation of $\ell(\by|\theta)$ computed using a standard bootstrap filter (BF) \cite{Gordon93,Doucet01} with $N$ particles (see Appendix F in \cite{Koblents14} for full details). One key feature of this approach, that we exploit for our analysis in Section \ref{sAnalysis}, is that $\ell^N(\by|\theta)$ can be proved to be an unbiased estimator of $\ell(\by|\theta)$ \cite{DelMoral04,Crisan15par}. 
 
The NPMC algorithm applied to the model described in Section \ref{sRepressilator}, with $K$ iterations, $M$ Monte Carlo samples per iteration, plain Gaussian proposals $\{ q_k \}_{k\ge1}$, and approximate likelihoods is outlined below.

\noindent \textit{\textbf{Initialisation}}. Draw $M$ i.i.d. samples $\theta_0^1, \theta_0^2, \ldots, \theta_0^M$ from the prior pdf $p_0(\theta)$. Then,
\begin{enumerate}
	\item compute non-normalised IWs $\tilde \sw_0^i \propto \ell^N(\by|\theta_0^i)$, $i=1, ..., M$, 
	\item compute TIWs as $\hat \sw_0^i = \mT_M\left(i, \{\tilde \sw_0^j\}_{j=1}^M\right)$, where $\mT_M : \{1, \ldots, M\} \times \{ \tilde \sw_0^j \}_{j=1}^M \rw [0, +\infty)$ is a nonlinear random map (note that the IWs $\tilde \sw_0^i$ are random),
	\item normalise the TIWs, $\sw_0^i = \frac{\hat \sw_0^i}{\sum_{j=1}^M \hat \sw_0^j}$, $i=1, ..., M$.
\end{enumerate}

\noindent \textit{\textbf{Iterative step}}. For $k = 1, \ldots, K$, take the following steps:
\begin{enumerate}
\item Let $q_k(\theta)=N(\theta|\mu_k,\Sigma_k)$ be a multivariate Gaussian pdf with mean vector and covariance matrix obtained, respectively, as
$$
\mu_k = \sum_{i=1}^M \sw_{k-1}^i \theta_{k-1}^i
\quad \mbox{and} \quad 
\Sigma_k = \sum_{i=1}^M \sw_{k-1}^i \left(
	\theta_{k-1}^i - \mu_k
\right)\left(
	\theta_{k-1}^i - \mu_k
\right)^\top.
$$
Note that the random variates $\theta_{k-1}^i$, $i=1, ..., M$, are $4 \times 1$ vectors. 

\item Draw $\theta_k^i$, $i=1, ..., M$, i.i.d. samples from $q_k(\theta)$.
\item Compute IWs, $\tilde \sw_k^i = \frac{ \ell^N(\by|\theta_k^i) p_0(\theta_k^i) }{ q_k(\theta_k^i) }$, $i=1, ..., M$.
\item Compute TIWs, $\hat \sw_k^i = \mT_M\left( i,\{\tilde \sw_k^j\}_{j=1}^M \right)$, $i=1, ..., M$, using the same nonlinear map as for $k=0$.
\item Normalise the TIWs, $\sw_k^i = \frac{\hat \sw_k^i}{\sum_{j=1}^M \hat \sw_k^j}$, $i=1, ..., M$.
\end{enumerate}

Following \cite{Koblents14}, the nonlinear map $\mT_M$ of choice is a ``clipping'' transformation. In particular, let $i_1, i_2, ..., i_M$ be a permutation of the indices $1, 2, ..., M$ such that the IWs become ordered, namely $\tilde \sw_k^{i_1} \ge \tilde \sw_k^{i_2} \ge \cdots \ge \tilde \sw_k^{i_M}$. The clipping transformation $\mT_M$, with parameter $1 \le M_c \le \sqrt{M}$, flattens the $M_c$ largest IWs and makes them equal to the $M_c$-th non-normalised IW, $\tilde \sw_k^{i_{M_c}}$. Specifically, for each $i=1, ..., M$, we obtain
\begin{equation}
\hat \sw_k^j = \mT_M\left( j,\{\tilde \sw_k^l\}_{l=1}^M \right) = \left\{
	\begin{array}{ll}
		\tilde \sw_k^{i_{M_c}}, &\mbox{if $\tilde \sw_k^j \ge \tilde \sw_k^{M_c}$},\\
		\tilde \sw_k^j, &\mbox{if $\tilde \sw_k^j < \tilde \sw_k^{M_c}$},\\
	\end{array}
\right..
\label{eqClipping}
\end{equation}
Other choices of $\mT_M$ are possible (e.g., tempering schemes \cite{Koblents14}).

Let $p_\by(\sd\theta)=p(\theta|\by)\sd\by$ denote the posterior probability measure (conditional on the observed data $\by$) of the parameter vector $\theta$. This measure yields the full probabilistic description of $\theta$ given the available observations. If  $p_\by(\sd\theta)$ is available, then we can compute various types of estimators and assess the associated errors. For example, the posterior-mean estimator is 
$$
\hat \theta_* = \int \mbI_\sS(\theta) \theta p_\by(\sd\theta),
$$ 
where 
$$
\mbI_A(z) = \left\{
	\begin{array}{cc}
	1, &\mbox{if $x \in A$},\\
	0, &\mbox{otherwise.}\\
	\end{array}
\right.
$$
This estimator minimises the mean square error (MSE), which, for an arbitrary estimator $\tilde \theta$, can also be written as an integral w.r.t. $p_\by(\sd\theta)$, namely, 
$$
\mbox{MSE}(\hat \theta) = \int \mbI_\sS(\theta) (\theta - \hat \theta)^2 p_\by(\sd\theta).
$$

The proposed NPMC algorithm yields a sequence of importance sampling (i.e., weighted Monte Carlo) approximations of $p_\by(\sd\theta)$. To be specific, at each iteration $k$ we obtain the random probability measure
$$
p_{\by,k}^M(\sd\theta) = \sum_{i=1}^M \sw_k^i \delta_{\theta_k^i}(\sd\theta),
$$ 
where $\delta_{\theta_k^i}$ denotes the Dirac delta measure centred at $\theta_k^i$. Using $p_{\by,k}^M(\sd\theta)$ we can approximate any parameter estimator. For instance,  
$
\hat \theta_* \approx \hat \theta^M_k = \sum_{i=1}^M \sw_k^i \theta_k^i
$ 
is the approximation of the posterior mean $\hat \theta_*$. The corresponding minimum MSE can also be approximately computed as
$$
\mbox{MSE}(\hat \theta^M_k) = \sum_{i=1}^M \sw_k^i \| \theta_k^i - \hat \theta^M_k \|^2.
$$

\begin{nota}
The statement of NPMC algorithm in this section is fairly general. It can be applied, with minor adjustments, to a very broad class of problems involving the Bayesian estimation of unknown parameters in state space stochastic models. 
\end{nota}

%
\section{Analysis} \label{sAnalysis} \label{sConvergence}

Consider a single iteration $k$ in the NPMC algorithm, with a fixed importance density $q_k \equiv q$. We refer to the random measure $p_{\by,k}^M(\sd\theta) = \sum_{i=1}^M \sw_k^i \delta_{\theta_k^i}(\sd\theta)$ computed via the TIWs $\sw_k^i$, $i=1, .., N$, as a nonlinear importance sampling (NIS) approximation of $p_\by(\sd\theta)$. Our aim in this section is to assess whether $p_{\by,k}^M(\sd\theta)$ converges towards the true measure $p_\by(\sd\by)$ or not as $M\rw\infty$. To do this, here are two issues that need to be handled and make the analysis more difficult compared to a conventional IS method (that relies on the standard IWs, rather than the TIWs). These issues are:
\begin{itemize}
\item[(i)] the distortion in the Monte Carlo approximation due to the clipping of the weights, which can be expected to introduce some bias (compared to the use of standard IWs); and
\item[(ii)] the impossibility to compute the IWs, and hence the TIWs, exactly, since the likelihood $\ell(\by|\theta)$ is intractable and we work with a particle approximation  $\ell^N(\by|\theta)$ instead.
\end{itemize} 
In \cite{Koblents14} it was proved that, when the IWs can be computed exactly, the NIS approximation converges almost surely (a.s.) towards the target probability measure as $M \rw \infty$, which accounts for (i) above\footnote{The analysis of  \cite{Koblents14}  does not provide convergence rates, though.}. The problem of the approximate computation of the weights was partially addressed in \cite{Koblents16}, for a relatively simple case where the errors in the IWs where assumed deterministic and bounded. However, the estimation problem studied in \cite{Koblents16} (parameter estimation for $\alpha$-stable distributions using iid data) did not involve any dynamics and the converge analysis only showed an upper bound for the approximation errors that included a deterministic constant (non-vanishing) term proportional to the IW approximation error.

Here, we show stronger analytical results that ensure the a.s. convergence of the NIS approximation when $M\rw\infty$ and can only be estimated as $\ell^N(\by|\theta)$, i.e., using a bootstrap filter with a finite and fixed number of particles $N$. Under assumptions which are standard in the classical IS theory, we prove that integrals of the form $\int f(\theta) p_{\by,k}^M(\sd\theta)$ converge towards $\int f(\theta) p_{\by,k}(\sd\theta)$ a.s. as $M\rw\infty$ and provide explicit convergence rates.

\subsection{Notation}

Since we focus our attention in the NIS scheme alone, i.e., a single iteration of the proposed algorithm, in the remaining of this section we drop the iteration index $k$. Hence, we assume a fixed importance density $q(\theta)$, from where $M$ independent (yet not necessarily identically distributed) Monte Carlo samples, $\theta^1, \theta^2, \ldots, \theta^M$, are drawn. Since the observations $\by$ are assumed arbitrary but fixed, we drop them from the likelihood notation and write 
$$
\ell(\theta)\dfn\ell(\by|\theta) \quad \mbox{and} \quad \ell^N(\theta)\dfn\ell^N(\by|\theta).
$$  
Then, the non-normalised IWs are approximated as 
$$
\tilde \sw^i = g^N(\theta^i) \dfn \frac{\ell(\theta^i)p_0(\theta^i)}{q(\theta^i)},
$$
where we have introduced the weight function $g^N=\ell^N p_0 / q$ as a shorthand. This weight function is a random approximation of the deterministic function $g=\ell p_0/q$. The support of $g$ is the same as the support of $q$, $\ell$ and $p_0$, denoted $\sS \subseteq \mathbb{R}^4$. We assume that $g(\theta) > 0$ for every $\theta \in \sS$ as well (a standard assumption in classical IS). It is also apparent that $p_\by \propto gq$, with the proportionality constant independent of $\theta$.

The non-normalised TIWs computed via the clipping function \eqref{eqClipping} are denoted 
$$
\hat \sw^i = [\mT^M \circ g^N](\theta^i) ),
$$
where $\circ$ represents function composition and we omit the index argument of \eqref{eqClipping} for conciseness (its value is clear from the notation in any case). The normalised TIWs are $\sw^i = \frac{\hat \sw^i}{\sum_{j=1}^M \hat \sw^j}$, and they are used to compute the approximate measure $p_\by^M(\sd\theta)$.
 
\subsection{Assumptions and a preliminary result}

Let the state sequence $\bx_n$ take values on some set $\mX \subseteq \Real^{14}$. We make the following assumptions on the conditional pdf of the observations $\by_n$, $n=1, 2, \ldots, R$, the prior density of the parameters, $p_0(\theta)$, and the importance function $q(\theta)$. 

\begin{hipotesis} \label{asBounds_on_g}
The observation sequence $\by_{1:R}$ is arbitrary but fixed. The functions $l_n(\by_n|\cdot):\mX\rw(0,\infty)$, $n=1, 2, ..., R$, are uniformly upper bounded and bounded away from zero, i.e., there exist finite and positive constants
$$
\| l \|_\infty = \sup_{n \ge 1, \bx_n \in \mX} l_n(\by_n|\bx_n) < \infty
\quad \mbox{and} \quad
l_{\sf inf} = \inf_{n \ge 1, \bx_n \in \mX} l_n(\by_n|\bx_n) > 0
$$
\end{hipotesis}
\begin{hipotesis} \label{asBounds_on_pq}
The ratio of pdf's $\frac{p_0(\theta)}{q(\theta)}$ is upper bounded and bounded away from zero on $\sS$, i.e., there exist positive and finite constants
$$
\left\| 
	\frac{p_0}{q}
\right\|_\infty = \sup_{\theta \in \sS} \left|
	\frac{
		p_0(\theta)
	}{
		q(\theta)
	}
\right| < \infty 
\quad \mbox{and} \quad
\left( \frac{p_0}{q} \right)_{\sf inf} = \inf_{\theta \in \sS} \frac{p_0(\theta)}{q_0(\theta)} > 0.
$$
\end{hipotesis}

\begin{nota}
If the parameter support set $\sS$ is compact, then A.\ref{asBounds_on_g} and A. \ref{asBounds_on_pq} hold naturally for most models of practical interest.
\end{nota}

The following lemma plays a key role in the asymptotic convergence analysis of the approximation $p_\by^M(\sd\theta)$. It states that $\ell^N(\theta)$ is an unbiased estimator of the likelihood $\ell(\theta)$ and enables us to show that the NIS scheme converges when $M\rw\infty$, even if the number of particles $N$ in the approximation $\ell^N(\theta)$ remains constant. 
\begin{lema} \label{lmUnbiased}
If Assumption \ref{asBounds_on_g} holds then 
$$
\max \{ \ell(\theta), \ell^N(\theta) \} \le \| l \|_\infty^R < \infty, \quad
\min\{ \ell(\theta), \ell^N(\theta) \} \ge \l_{\sf inf}^R \quad 
\quad \mbox{and} \quad
E\left[
	\ell^N(\theta)
\right] = \ell(\theta)
$$
independently of $N$. 
\end{lema}
\noindent\textbf{Proof.} From the definition of $\ell(\theta)$ in Eq. \eqref{eqDefEll} and its estimator $\ell^N(\theta)$ in Eq. \eqref{eqUBEstimator}, it is clear that both $\ell(\theta) \le \| l \|_\infty^R$ and $\ell^N(\theta) \le \| l \|_\infty^R$ when $R$ is the number of available observations. It is also straightforward to show that $\ell(\theta) \ge l_{\sf inf}^R$ and $\ell^N(\theta) \ge l_{\sf inf}^R$. The fact that $\ell^N(\theta)$ is unbiased is a consequence of \cite[Theorem 7.4.2]{DelMoral04} (see also \cite[Lemma 2]{Crisan15par} for an alternative proof that does not rely on the Feynmann-Kac framework). \qed

\subsection{Asymptotic convergence}

In the sequel we look into the approximation of integrals of the form 
$$
(f,p_\by) \dfn \int \mbI_{\sS}(\theta) f(\theta) p_\by(\sd\theta),
$$ 
where $\mbI_\sS(\theta)$ is an indicator function (namely, $\mbI_\sS(\theta)=1$ if $\theta \in \sS$ and $\mbI_\sS(\theta)=0$ otherwise) and $f$ is a bounded real function in the parameter space $\sS$. We use $\|f\|_\infty \dfn \sup_{\bftheta \in \sS}|f(\bftheta)| < \infty$ to denote the supremum norm of a bounded function while the set of bounded functions on $\sS$ is denoted $B(\sS)$. The approximations of interest are
$$
(f,p_\by) \approx (f,p_\by^M) = \sum_{i=1}^M f(\theta^i) \sw^i,
$$
for any $f \in B(\sS)$.

The following Theorem yields an explicit upper bound for the (random) approximation error $| (f,p_\by^M) - (f,p_\by) |$. The bound is proportional to $M^{-\frac{1}{2}}$ and, therefore, it vanishes as $M \rw \infty$, independently of the number of particles $N$ used in the approximation of the likelihoods $\ell^N(\theta^i)$.

\begin{teorema} \label{thBasic}
Assume that A.\ref{asBounds_on_g} and A.\ref{asBounds_on_pq} hold and $M_c \le \sqrt{M}$. Then, for every $\epsilon \in \left(0,\frac{1}{2}\right)$ (arbitrarily small) and every $f \in B(\sS)$ there exists a positive and a.s. finite random variable $V_{f,\epsilon}$, independent of $M$ and $M_c$, such that
\begin{equation}
| (f,p_\by^M) - (f,p_\by) | \le \frac{
    V_{f,\epsilon}
}{
    M^{\frac{1}{2}-\epsilon}
}.
\nonumber
\end{equation}
\end{teorema}

\noindent\textbf{Proof.} Recall the intractable weight function $g=\ell p_0 / q$ and its random estimator $g^N = \ell^N p_0 / q$. The integral of any $f \in B(\sS)$ w.r.t. the posterior measure $p_\by(\sd\theta) \propto \ell(\theta)p_0(\theta)\sd\theta$ can be written as
\begin{equation}
(f,\pi) = \frac{
    (fg,q)
}{
    (g,q)
}
\nonumber
\end{equation}
by simply noting that $g(\theta)q(\theta)\sd\theta = \ell(\theta)p_0(\theta)\sd\theta$. Similarly, we can construct an approximation $\hat p_\by^M(\sd\theta) = \frac{1}{\sum_{j=1}^M g^N(\theta^j)} \sum_{i=1}^M g^N(\theta^i) \delta_{\theta^i}(\sd\theta)$ of the posterior $p_\by(\sd\theta)$ using the IWs {\em before} the clipping, and then obtain the approximate integral
\begin{equation}
(f,\hat p_\by^M) = \frac{
    (fg^N,q^M)
}{
    (g^N,q^M)
}
\label{eqQQ0}
\end{equation}
where $q^M(\sd\theta) = \frac{1}{M} \sum_{i=1}^M \delta_{\theta^i}(\sd\theta)$. It is simple to show that
\begin{equation}
(f,\hat p_\by^M) - (f,p_\by) = \frac{
    (fg^N,q^M) - (fg,q)
}{
    (g,q)
} + (f,p_\by^M) \frac{
    (g,q) - (g^N,q^M)
}{
    (g,q)
} \label{eqDecom}
\end{equation}
and, since $(g,q)=(\ell,p_0) = \int \mbI_\sS(\theta)\ell(\theta)p_0(\theta)\sd\theta > 0$ and $(f,\hat p_\by^M) \le \| f
\|_\infty$, Eq. \eqref{eqDecom} readily yields
\begin{equation}
| (f,\hat p_\by^M) - (f,p_\by) | \le \frac{
    1
}{
    (\ell,p_0)
} \left|
    (fg^N,q^M) - (fg,q)
\right| + \frac{
    \| f \|_\infty
}{
    (\ell,p_0)
} \left|
    (g,q) - (g^N,q^M)
\right|, \label{eqInicial}
\end{equation}
and, therefore, the problem of computing a bound for $| (f,\hat p_\by^M) - (f,p_\by) |$ reduces to computing bounds for errors of the form $| (bg^N,q^M) - (bg,q) |$, where $b \in B(\sS)$.

Choose any $b \in B(\sS)$. A simple triangle inequality yields
\begin{equation}
| (bg^N,q^M) - (bg,q) | \le | (bg^N,q^M) - (bg,q^M) | + | (bg,q^M) - (bg,q) |. \label{eqTriang1}
\end{equation}
Since $q^M = \frac{1}{M} \sum_{i=1}^M \delta_{\theta^i}$, for the second term on the right hand side of \eqref{eqTriang1} we can write
\begin{equation}
\mbE\left[
    | (bg,q^M) - (bg,q) |^p
\right] = \mbE\left[
    \left|
        \frac{1}{M} \sum_{i=1}^M Z^i
    \right|^p
\right], \label{eqTriang2t}
\end{equation}
where the random variables
\begin{equation}
Z^i = b(\theta^i)g(\theta^i) - (bg,q), \quad i=1, ..., M, \nonumber
\end{equation}
are independent, with zero mean (recall the $\theta^{(i)}$'s are i.i.d. draws from $q$) and bounded, because $b$ is bounded and A.\ref{asBounds_on_g} and A.\ref{asBounds_on_pq} imply that $g<\| l \|_\infty^R \times \left\|  \frac{p_0}{q} \right\|_\infty < \infty$. Therefore, it is an exercise in combinatorics to show that
\begin{equation}
\mbE\left[
    \left|
        \frac{1}{M} \sum_{i=1}^M Z^{(i)}
    \right|^p
\right] \le \frac{
    \tilde c^p \| l \|_\infty^{Rp} \left\|  \frac{p_0}{q} \right\|_\infty^p \| b \|_\infty^p
}{
    M^\frac{p}{2}
}, \label{eqZygmund}
\end{equation}
where $\tilde c$ is a constant independent of $M$ and $q$. Combining \eqref{eqZygmund} with \eqref{eqTriang2t} readily yields
\begin{equation}
\| (bg,q^M) - (bg,q) \|_p \le \frac{
    \tilde c \| l \|_\infty^R \left\|  \frac{p_0}{q} \right\|_\infty \| b \|_\infty
}{
    \sqrt{M}
}. \label{eq_tri2term}
\end{equation}
The inequality \eqref{eq_tri2term} implies that there exists an a.s. finite random variable $\tilde U_{b,\epsilon}>0$ such that
\begin{equation}
| (bg,q^M) - (bg,q) | \le \frac{
    \tilde U_{b,\epsilon}
}{
    M^{\frac{1}{2}-\epsilon}
}, \label{eq_tri2term_eps}
\end{equation}
where $0 < \epsilon < \frac{1}{2}$ is an arbitrarily small constant independent of $M$ (see \cite[Lemma 4.1]{Crisan14a}).

If we expand the first term on the right hand side of \eqref{eqTriang1} we arrive at
\begin{eqnarray}
\left|
    (bg^N,q^M) - (bg,q^M)
\right| &=& \left|
    \frac{1}{M} \sum_{i=1}^M b(\theta^i) \left(
    		g^N(\theta^i) - g(\theta^i)
	\right)
\right| \nonumber \\
&=& \left|
    \frac{1}{M} \sum_{i=1}^M Z_N^i
\right|, \label{eqZ2}
\end{eqnarray}
where the r.v.'s $Z_N^i = \frac{b(\theta^i)p_0(\theta^i)}{q(\theta^i)}\left(  \ell^N(\theta^i) - \ell(\theta^i) \right)$, $i=1, 2, ..., M$, are independent (because the samples $\theta^1, \ldots, \theta^M$ are independent) and zero mean, as a result of Lemma \ref{lmUnbiased}. Since they are also bounded, namely $| Z_N^i | \le \| b \|_\infty \| l \|_\infty^R \left\| \frac{p_0}{q}\right\|_\infty$ as a consequence of A.\ref{asBounds_on_g} and A.\ref{asBounds_on_pq}, it is again an exercise to show that \eqref{eqZ2} implies
\begin{equation}
E\left[
	\left|
    		(bg^N,q^M) - (bg,q^M)
	\right|^p
\right] \le \frac{
    \bar c^p \| l \|_\infty^{Rp} \left\|  \frac{p_0}{q} \right\|_\infty^p \| b \|_\infty^p
}{
    M^\frac{p}{2}
}
\label{eqZygmund2}
\end{equation}
in the same manner as we obtained the inequality \eqref{eqZygmund}. Resorting again to \cite[Lemma 4.1]{Crisan14a}, from \eqref{eqZygmund2} we deduce that there exists an a.s. finite random variable $\bar U_{b,\epsilon}>0$, independent of $M$, such that
\begin{equation}
| (bg^N,q^M) - (bg,q^M) | \le \frac{
    \bar U_{b,\epsilon}
}{
    M^{\frac{1}{2}-\epsilon}
}, \label{eq_tri1term_eps}
\end{equation}
where $0 < \epsilon < \frac{1}{2}$ is an arbitrarily small constant independent of $M$.
 
Taking together \eqref{eqTriang1}, \eqref{eq_tri2term_eps} and \eqref{eq_tri1term_eps} we arrive at
\begin{equation}
| (bg^N,q^M) - (bg,q) | \le \frac{
    U_{b,\epsilon}
}{
    M^{\frac{1}{2}-\epsilon}
}, \label{eqBasica}
\end{equation}
where $U_{b,\epsilon}=\tilde U_{b,\epsilon} + \bar U_{b,\epsilon} \ge 0$ is an a.s. finite r.v. independent of $M$, and $\epsilon \in \left(0,\frac{1}{2}\right)$ can be chosen to be arbitrarily small.   

It is now immediate to combine the inequality \eqref{eqInicial} with the bound in \eqref{eqBasica}. If we choose  $b=f$ in order to control the first term on the right hand side of \eqref{eqInicial}, and $b=1$ in order to control the second term, we readily find that
\begin{equation}
| (f,\hat p_\by^M) - (f,p_\by) | \le \frac{
	\tilde V_{f,\epsilon}
}{
	M^{\frac{1}{2}-\epsilon}
}, \label{eqProof_ineq1}
\end{equation}
where
$$
\tilde V_{f,\epsilon} = \frac{ U_{f,\upsilon} + \| f \|_\infty U_{1,\epsilon}}{(\ell,p_0)} > 0
$$
is an a.s. finite random variable independent of $M$, and $\epsilon \in \left(0,\frac{1}{2}\right)$ can be chosen arbitrarily small. 

Having found the bound in \eqref{eqProof_ineq1}, it is now relatively straightforward to compute the desired error rate for $| (f,p_\by^M)-(f,p_\by) |$. We first apply a triangle inequality to obtain
\begin{equation}
| (f,p_\by^M) - (f,p_\by) | \le 
| (f,p_\by^M) - (f,\hat p_\by^M) | + | (f,\hat p_\by^M) - (f,p_\by) | \label{eqTriang2}
\end{equation}
and \eqref{eqProof_ineq1} directly yields a bound for the second term on the right hand side of \eqref{eqTriang2}. For the first term, we note that
\begin{equation}
(f,p_\by^M) = \frac{
    ( f[\varphi^M \circ g^N], q^M )
}{
    ( \varphi^M \circ g^N, q^M )
}, \label{eqNormalizada}
\end{equation}
where $\circ$ denotes composition, hence $(\varphi^M \circ g^N)(\bftheta) = \varphi^M( g^N (\bftheta) )$. If we now use \eqref{eqNormalizada} and the expression for $(f,\hat p_\by^M)$ in \eqref{eqQQ0} we obtain, by the same argument leading to \eqref{eqInicial}, that
\begin{eqnarray}
| (f,p_\by^M) - (f,\hat p_\by^M) | &\le& \frac{
    | (f[\varphi^M\circ g^N], q^M) - (fg^N,q^M) |
}{
    (\varphi^M \circ g^N, q^M)
} 
+ \frac{
    \| f \|_\infty | (\varphi^M\circ g^N, q^M) - (g^N,q^M) |
}{
    (\varphi^M \circ g^N, q^M)
}  \nonumber \\
&\le& l_{\sf inf}^{-R} \left(\frac{p_0}{q}\right)_{\sf inf} | (f[\varphi^M\circ g^N], q^M) - (fg^N,q^M) | \nonumber\\
&& + l_{\sf inf}^{-R} \left(\frac{p_0}{q}\right)_{\sf inf}  \| f \|_\infty | (\varphi^M\circ g^N, q^M) - (g^N,q^M) |, \nonumber\\
\label{eqTriang3}
\end{eqnarray}
where the second inequality follows from the definition of the clipping transformation $\varphi^M$ and the bound $g^N \ge l_{\sf inf}^{-R} \left(\frac{p_0}{q}\right)_{\sf inf}$. The latter bound is easily obtained by combining A.\ref{asBounds_on_pq} and Lemma \ref{lmUnbiased}.

In order to use \eqref{eqTriang3}, we look into errors of the form $|(b[\varphi^M\circ g^N], q^M) - (bg^N,q^M)|$ for arbitrary $b \in B(\sS)$. This turns out relatively straightforward since, from the construction of $\varphi^M$,
\begin{equation}
|(b[\varphi^M\circ g^N], q^M) - (bg^N,q^M)| =
\left|
    \frac{1}{M}\sum_{r=1}^{M_c} b(\theta^{i_r}) \left[
        g^N(\theta^{i_{M_c}}) - g^N(\bftheta^{i_r})
    \right]
\right| \le 2 \| l \|_\infty^R \left\| \frac{p_0}{q} \right\|_\infty \|b\|_\infty \frac{M_c}{M},\nonumber\\
\label{eqFacil}
\end{equation}
where the inequality follows from the bound $g^N \leq \| l \|_\infty^R \left\| \frac{p_0}{q} \right\|_\infty$, which is a straightforward consequence of assumptions A.\ref{asBounds_on_g} and A.\ref{asBounds_on_pq} and the definition of the estimate $g^N$. We can plug \eqref{eqFacil} into \eqref{eqTriang3} twice, first choosing $b=f$ and then $b=1$, in order to control the two terms in the triangle inequality. As a result, we arrive at the {\em deterministic} bound
\begin{equation}
| (f,p_\by^M) - (f,\hat p_\by^M) | \le \frac{
    4 \| l \|_\infty^R l_{\sf inf}^{-R} \left\| \frac{p_0}{q} \right\|_\infty \left( \frac{p_0}{q} \right)_{\sf inf} \|f\|_\infty M_c
}{
    M
} \le \frac{
    4 \| l \|_\infty^R l_{\sf inf}^{-R} \left\| \frac{p_0}{q} \right\|_\infty \left( \frac{p_0}{q} \right)_{\sf inf} \|f\|_\infty
}{
    \sqrt{M}
}, \label{eqSegunda}
\end{equation}
where the second inequality follows from the assumption $M_c\le\sqrt{M}$ in the statement of Theorem \ref{thBasic}.

Plugging \eqref{eqSegunda} and \eqref{eqProof_ineq1} into \eqref{eqTriang2} yields the desired error rate,
\begin{equation}
| (f,p_\by^M) - (f,p_\by) | \le \frac{
    V_{f,\epsilon}
}{
    M^{\frac{1}{2}-\epsilon}
},
\end{equation}
where 
$$
V_{f,\epsilon} = \tilde V_{f,\epsilon} + 4 \| l \|_\infty^R l_{\sf inf}^{-R} \left\| \frac{p_0}{q} \right\|_\infty \left( \frac{p_0}{q} \right)_{\sf inf} \|f\|_\infty
$$
is an a.s. finite random variable and $\epsilon < \frac{1}{2}$ is an arbitrarily small constant, both of them independent of $M$. \qed

\begin{nota}
Theorem 1 is a general result regarding nonlinear importance sampling. It holds true for any problem involving the approximation of the posterior probability distribution of the unknown parameters of a state space model as long as Assumptions 1 and 2 hold. It is, therefore, not uniquely linked to the repressilator model of interest in this paper.
\end{nota}
%
\section{Computer simulations} \label{sSimulations}

We have carried out computer simulations to  assess both the proposed model, i.e., the stochastic modified repressilator model described by Eqs.  \eqref{8eq}-\eqref{14eq}, and the NPMC parameter estimation algorithm of Section \ref{sNPMC}.

For all the simulations presented here, the model parameters are set to their standard values  (see Section \ref{stoch}) in order to generate synthetic (i.e., simulated) trajectories for the dynamic variables, ${ \bar\bx}_{i,m}$, $i=1,2$ and $m=0,1,\ldots,$ and sequences of synthetic observationns $\by_n$, $n=1,2,\ldots,$ according to Eq.\eqref{eqObserv}. This choice of parameters yields an underlying  chaotic (and noisy) dynamics of the state variables, as will be shown below. The integration step of the Euler-Mayurama scheme is $\sfh=10^{-3}$ time units. When needed, observations are generated every $m_o=20$ discrete-time steps of model  \eqref{8eq}-\eqref{14eq}  (equivalently, every $m_o\sfh=0.02$ continuous time units). The observational noise   $\bfepsilon_n$ in Eq. \eqref{eqObserv} is assumed to be zero-mean Gaussian with identity covariance matrix 
${\bf I}_2=\left[
\begin{array}{cc}
1 & 0	\\
0 & 1	\\
\end{array}
\right]$, and the standard deviation parameter in the same Eq. \eqref{eqObserv}  is set to $\sigma_y=1$.

\begin{figure*}[htb]
\centering \subfigure[]
{\includegraphics[width=0.33\linewidth]{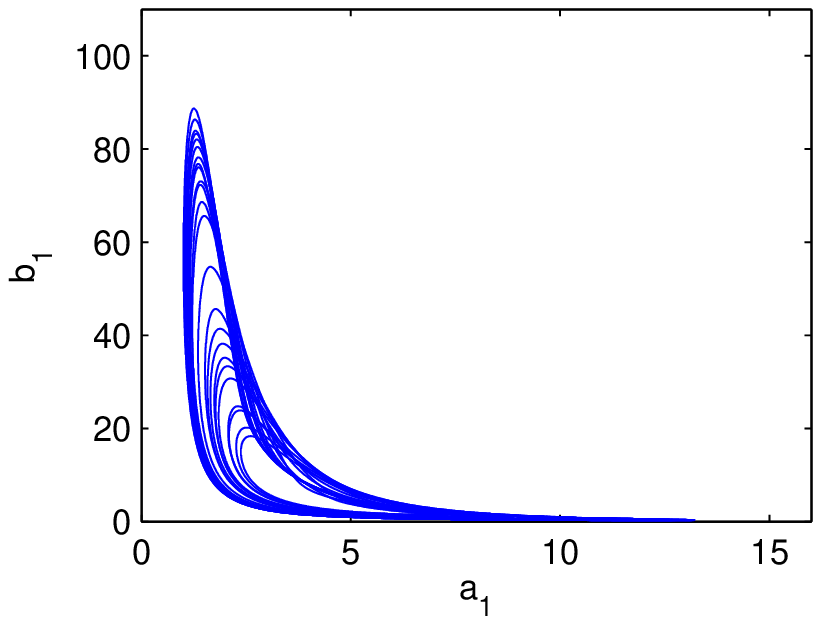}}
 \subfigure[]
{ \includegraphics[width=0.33\linewidth]{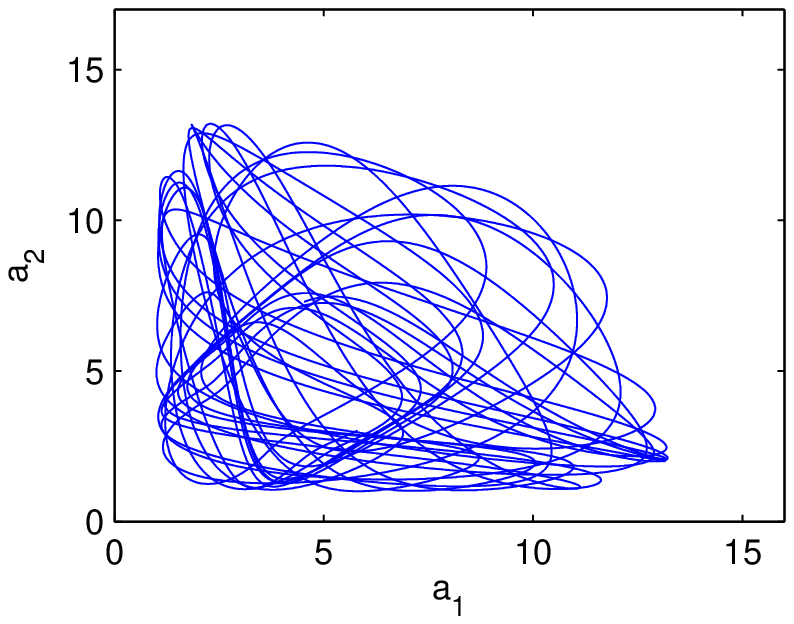}}
 \subfigure[]
{\includegraphics[width=0.33\linewidth]{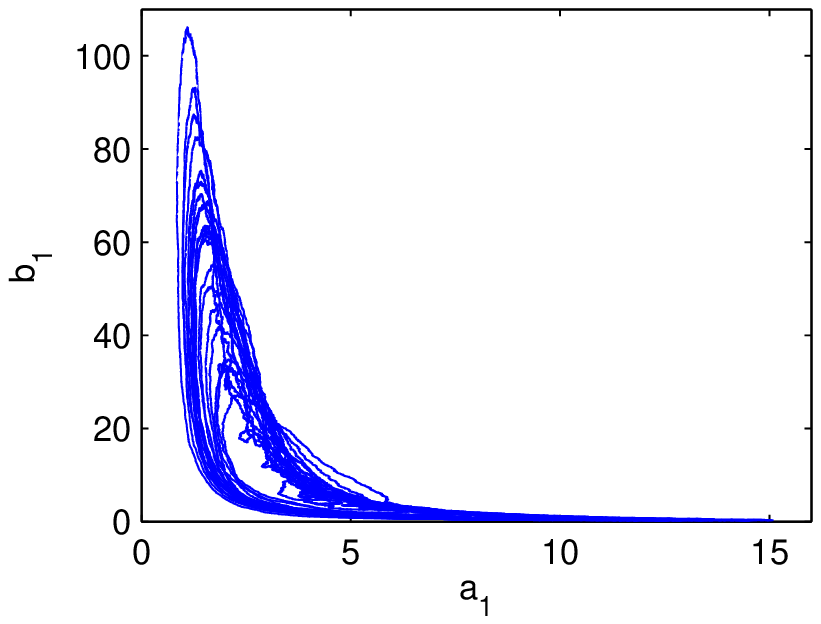}}
  \subfigure[]
 {\includegraphics[width=0.33\linewidth]{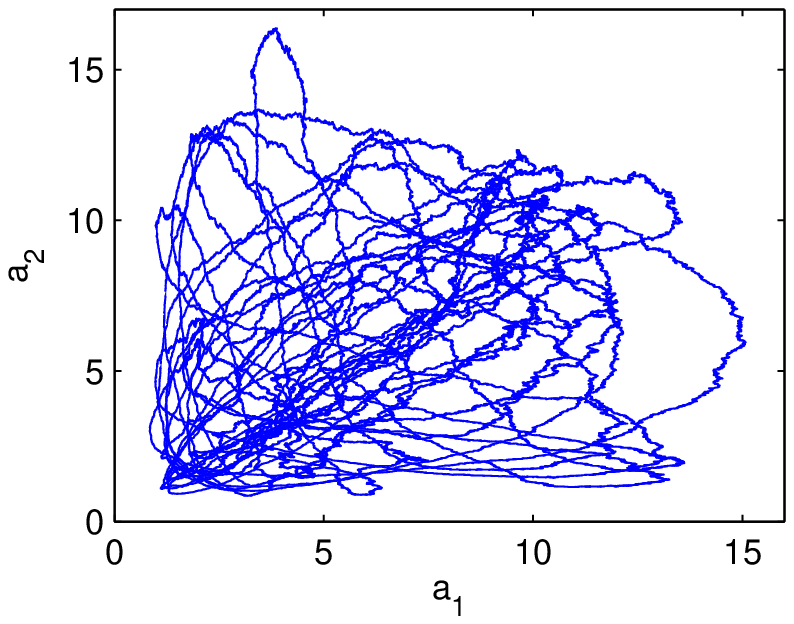}}
\caption{Comparison of 2-dimensional phase space diagrams  for the deterministic and the stochastic repressilator models (a) $b_1$ versus $a_1$ and (b) $a_2$ versus $a_1$ for  the determinist model; (c) $b_1$ versus $a_1$ and (d) $a_2$ versus $a_1$ for the stochastic model with  variance  $0.02^2$ for the dynamical noise.}
\label{determ_stoch}
\end{figure*}

When the parameters $\sigma_a$,  $\sigma_b$,  $\sigma_c$,  $\sigma_A$,  $\sigma_B$,  $\sigma_C$ and  $\sigma_S$, which control the variance of the process noise variables $w^{(j)}_{i,m}$, $j=1,\ldots,7$, $i=1,2$, are set to zero, we recover the deterministic modified repressilator of \cite{Marinho13}, which displays chaotic behaviour. For this specific case, we have run a long simulation of the noiseless system ($10000$ continuous time units) and used the results to obtain phase diagrams. In particular, Fig.\ref{determ_stoch}(a)  shows the phase diagram for the variable $b_1$ versus $a_1$, while Fig. \ref{determ_stoch}(b) depicts the phase diagram of $a_2$ versus $a_1$. What we observe are two views of the multidimensional chaotic attractor generated by this system.
When we add dynamical noise, by setting $\sigma_a^2=\sigma_b^2=\sigma_c^2=\sigma_A^2=\sigma_B^2=\sigma_C^2=\sigma_S^2=0.02^2$, we obtain an stochastic dynamical system. However, if we repeat the experiment to generate long trajectories (with the same initial conditions and the same duration) we obtain two similar phase diagrams, as shown in Figs. \ref{determ_stoch}(c) and  \ref{determ_stoch}(d). Indeed, the figures simply depict noisy (i.e., slightly perturbed) versions of the original deterministic attractor. This illustrates the fact that the underlying chaotic dynamics is preserved in the stochastic model, which can account for slight perturbations or uncertainties in the physical system as well.

From Figure \ref{determ_stoch} we also observe that the trajectories of the dynamic variables remain confined within a bounded region. This is relevant because if the state space $\mX\subset\Real^4$ can be ensured to be compact, then the assumptions A.\ref{asBounds_on_g} and A.\ref{asBounds_on_pq} on which Theorem \ref{thBasic} in Section \ref{sAnalysis} relies follow naturally.

In the sequel, we assess the performance of the proposed NPMC method and compare it with some alternative techniques that can be found in the literature. For the subsequent simulation experiments involving the NPMC algorithm, we simulate trajectories of the deterministic modified repressilator model (i.e., we use Eqs. \eqref{8eq}-\eqref{14eq} with $\sigma_a^2=\sigma_b^2=\sigma_c^2=\sigma_A^2=\sigma_B^2=\sigma_C^2=\sigma_S^2=0$) with random initial condition\footnote{The initial condition is generated from a multivariate Gaussian distribution with mean $(a_{1,0},b_{1,0},c_{1,0},A_{1,0},B_{1,0},C_{1,0},S_{1,0},a_{2,0},b_{2,0},c_{2,0},A_{2,0},B_{2,0},C_{2,0},S_{2,0})=(4.5,6,3,4.2,19,4.3,0.1,7.3,1.5,3.4,7,6.5,3.6,0.08)$ and covariant matrix $\sigma_0^2 {\bf I}$, where $\sigma_0^2=0.05^2$.} and then generate observations over an interval of $80$ continuous time units (hence $80/h=80\times10^3$ time steps in the Euler scheme). With observations collected every $m_o=20$ discrete steps, this yields a sequence of $4\times 10^3$ 2-dimensional observation vectors contaminated with zero mean Gaussian noise with unit variance. In order to compute the likelihood approximation $\ell^N(\theta)$, which is necessary to obtain IW's and TIW's, we use a bootstrap filter with $N=100$ particles.

\begin{figure*}[htb]
\centering \subfigure[]
	{\includegraphics[width=0.33\linewidth]{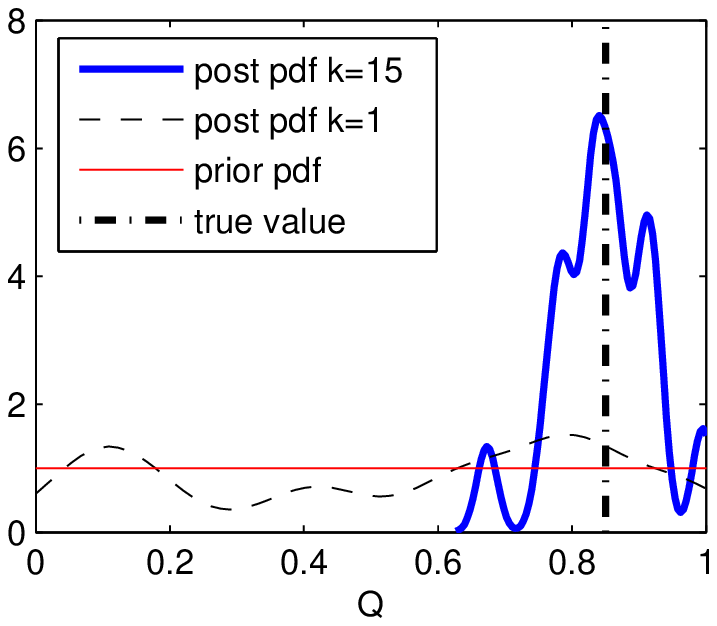}}
\subfigure[]
	{\includegraphics[width=0.34\linewidth]{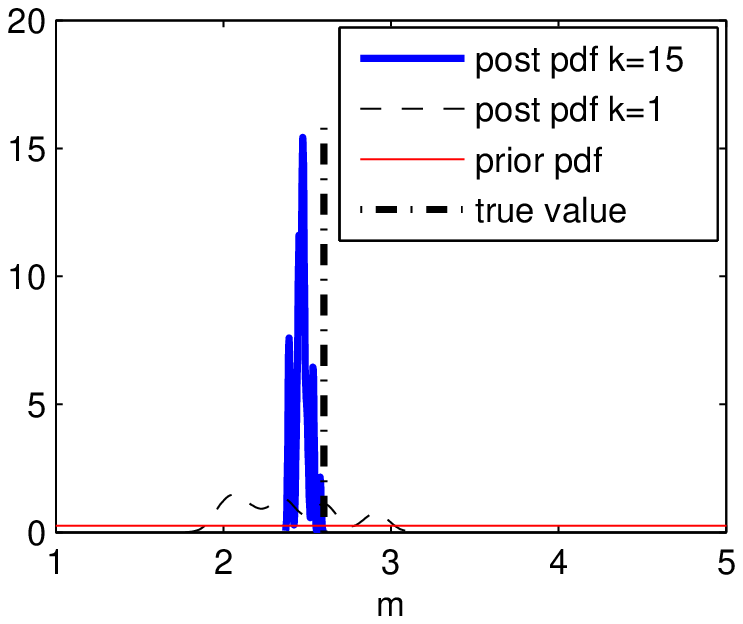}}
\subfigure[]	
	{\includegraphics[width=0.36\linewidth]{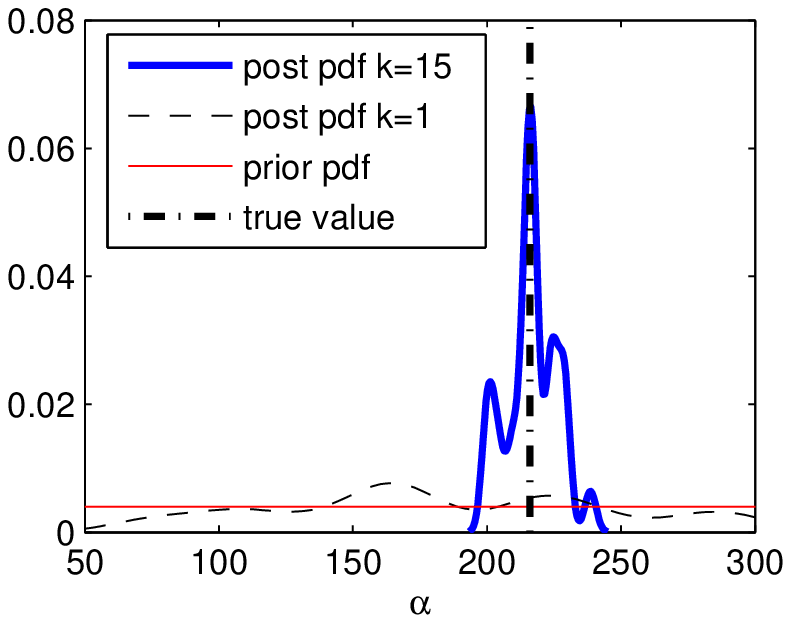}}
\subfigure[]	
	{\includegraphics[width=0.33\linewidth]{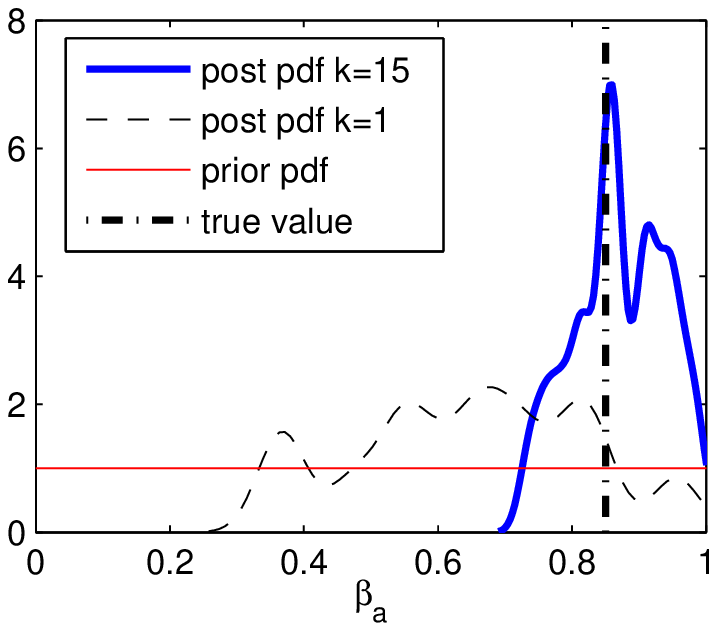}}
\caption{Posterior pdf's computed from the outcome of the NPMC algorithm for $M=200$ samples per iteration, compared with the true parameter values for the different parameters: (a) posterior density of $Q$, (b) posterior density of $\sfm$, (c) posterior density of $\alpha$, (d) posterior density of $\beta_a$. }
\label{fPDFs}
\end{figure*}

Figure  \ref{fPDFs} shows the outcome of a sample run of the NPMC algorithm applied to a sequence of $\frac{80}{m_oh}$ observations as described above, with a deterministic trajectory of the state variables and standard parameter values as ground truth. Each plot in Fig.  \ref{fPDFs}  depicts:
\begin{itemize}
\item the a priori uniform pdf (solid red line)
\item the approximate posterior pdf after one iteration of the NPMC algorithm, and
\item the approximate posterior pdf after $15$ iterations of the NPMC algorithm,
\end{itemize}
for an unknown model parameter (hence, for parameters $Q$,  $\sfm$,  $\alpha$ and $\beta_a$, from left to right and from top to bottom). Toghether with the densities, the actual parameter value is marked with a vertical dashed-dotted line. The NPMC algorithm was run with $M=200$ samples per iteration. The approximate pdf's are generated from these samples, $\{\theta_k^\}_{i=1}^M$, for iteration $k=1$ and $k=15$, using a Gaussian-kernel estimator.

From the plots, we observe that the probability mass of the approximations tends to concentrate around the region where the actual parameter value is located. In this simulation, the true values of  $Q$,   $\alpha$ and $\beta_a$ are well aligned with the main modes of the approximate pdf's , although this may not necessarily be the case for {\em all} simulations. Indeed, there is a bias between the actual value of $m$ and the main mode of the approximate density produced by the NPMC method. To explain this ``mismatch" let us recall that the approximate statistics generated by the algorithm converge to the true value of these statistics. For example, if we are interested in the posterior expected value of $\theta$, then we compute
\begin{equation}
{\hat \theta}_k^M =\sum_{i=1}^M w_k^i \theta_k^i
\end{equation}
and Theorem \ref{thBasic} ensures that 
\begin{equation}
\lim _{M \rightarrow \infty}{\hat \theta}_k^M =\int \theta p_{\bf y}({\bf \theta}) d{\bf \theta}=E[\theta|{\bf y}]
\quad \mbox{a.s.}
\end{equation}
However, depending on the available data (i.e., the dimension of vector ${\bf y}$), the posterior mean $E[\theta|{\bf y}]$ can be significantly different from the {\em true} value of $\theta$ used to generate the synthetic data.

\begin{figure*}[htb]
\centering \subfigure[]{
	\includegraphics[width=0.3\linewidth]{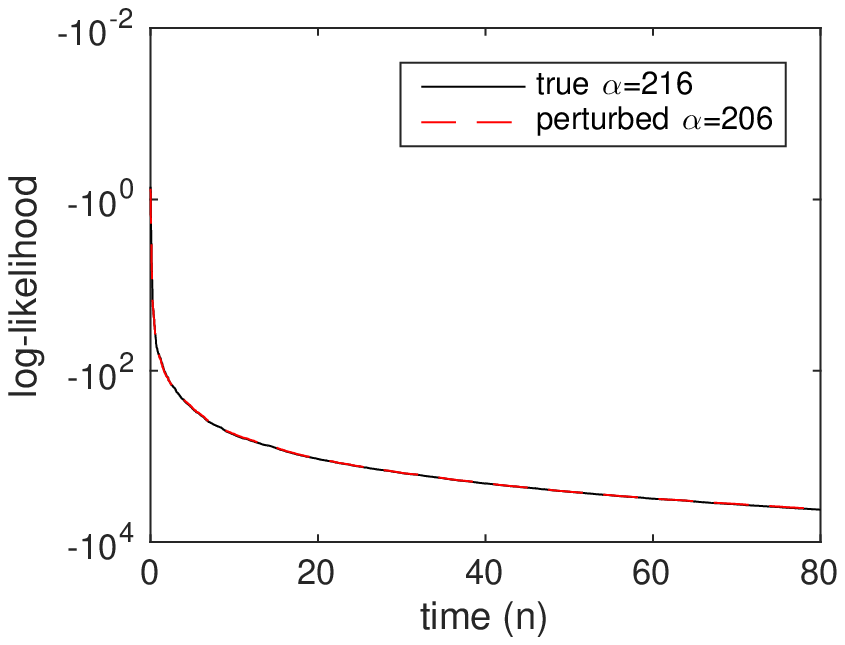}
}
 \subfigure[]{
 	\includegraphics[width=0.3\linewidth]{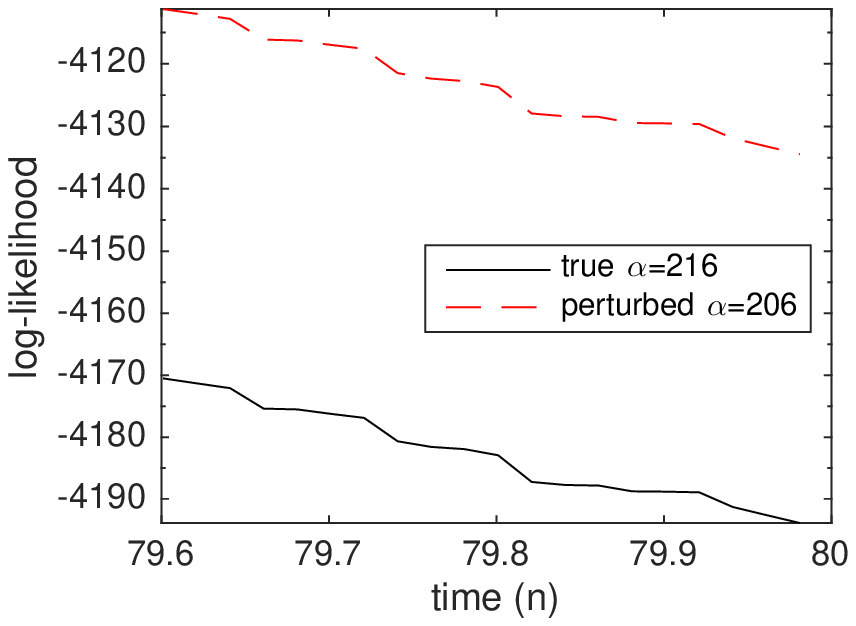}
}
 \subfigure[]{
 	\includegraphics[width=0.3\linewidth]{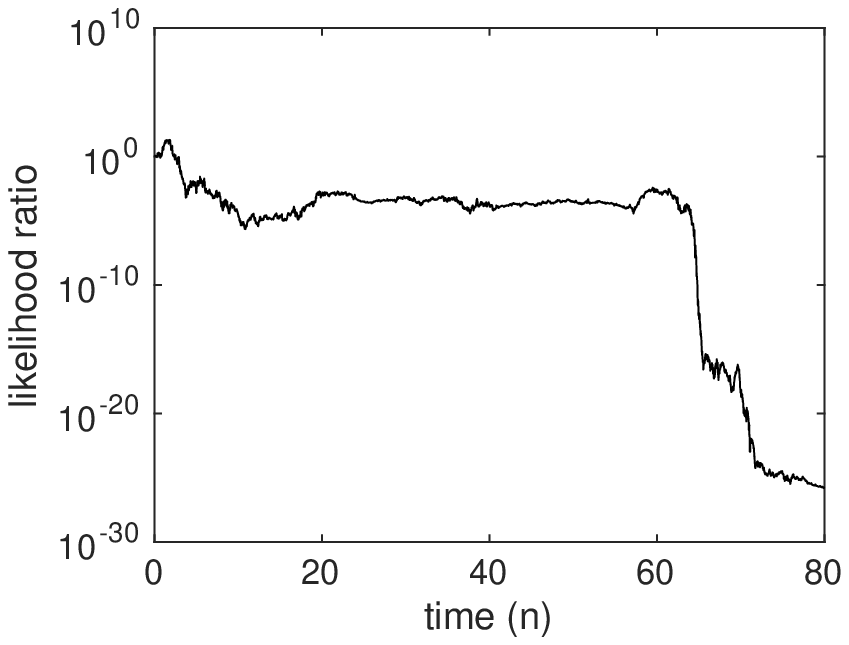}
}
\caption{Comparison of the approximate likelihood of the true parameter vector $\theta_*=[0.85,2.6,216,0.85]$ and perturbed version $\theta'=[0.85,2.6,\underline{206},0.85]$. (a) Approximate log-likelihoods, $\ell({\bf y}_{1:n}|\theta_*)$ and $\ell({\bf y}_{1:n}|\theta')$ versus time $n$. (b) Zoom of plot (a) for a shorter time interval, showing that the perturbed parameter $\theta'$ yields a higher likelihood for the observation sequence $\bf y$ generated in this computer experiment. (c) The likelihood ratio $\frac{\ell^N({\bf y}_{1:n}|\theta_*)}{\ell^N({\bf y}_{1:n}|\theta')}$ versus time $n$, showing that $\ell^N(\theta') > \ell^N(\theta_*)$.}
\label{fig4}
\end{figure*}

A simple way to illustrate this issue is to approximate the likelihood of two different parameter vectors, say ${\theta}_*=[Q_*,m_*,\alpha_*,\beta_{a*}]=[0.85,2.6,216,0.85]$ the ground truth, and $\theta'={\theta}_*+[0,0,-10,0]$ a mismatched version, and see that, for a common and fixed observation vector they are approximately the same (actually, $\ell^N(\theta')>\ell^N(\theta_*)$, even if the difference is small). This is shown in Fig. \ref{fig4}, which, for a fixed sequence ${\bf y}=\{ {\bf y}_1, {\bf y}_2,\ldots, {\bf y}_n,\ldots \}$, depicts the approximate log-likelihood $\log(\ell^N({\bf y}_{1:n}|\theta_*))$ and $\log(\ell^N({\bf y}_{1:n}|\theta'))$ versus $n$. The number of particles in the BF is set to $N=600$ in this case to ensure that we obtain low-variance estimates.

Finally, we have compared the performance of the proposed NPMC method with two other state-of-the-art techniques. The first one is the approximate Bayesian computation (ABC) sequential Monte Carlo (SMC) algorithm of \cite{Marinho13} while the second one is a particle Metropolis-Hastings (PMH)\cite{Andrieu10} method.

ABC SMC is a likelihood-free, Monte Carlo sampling algorithm that relies on a deterministic version of Eqs.  \eqref{8eq}-\eqref{14eq}  modified to include a coupling term that allows to drive the dynamical system using the observations. The ABC principle involves
\begin{itemize}
\item  drawing random candidates for the parameter vector, say $\theta^1,\theta^2,\ldots,$ 
\item using these candidates to generate synthetic data, say ${\bf y}(\theta^1), {\bf y}(\theta^2),\ldots,$ by means of the deterministic reppresilator dynamics 
\item and then comparing the actual and synthetic data using some suitable distance $d$, i.e., evaluating $d({\bf y},{\bf y}(\theta^1)),d({\bf y},{\bf y}(\theta^2)),\ldots$
\end{itemize}
Samples that yield a small enough distance, typically below a prescribed tolerance, $\epsilon$, are accepted, and those yielding large distances are discarded. Candidates are drawn and tried until a prescribed number of them are accepted. In a SMC setup, this procedure is repeated over several stages with decreasing tolerances $\epsilon_1>\epsilon_2>\ldots$ (see \cite{Marinho13} for details). Here we have applied the algorithm with tolerances $\epsilon_1=3$, $\epsilon_2=2.5$, $\epsilon_3=2.3$, $\epsilon_4=2.2$ and $\epsilon_5=2.1$ and the aim of accepting $15\times 400/5=1200$ samples per stage. The maximum number of Monte Carlo draws per stage, however, is set to $1600\times10^3$, to prevent the algorithm from getting stuck at any stage due to low acceptance rates.

The PMH algorithm is a Markov chain Monte Carlo (MCMC) technique\cite{Robert04}. It generates candidates with a Gaussian kernel with covariance matrix 
$${\bf \Sigma}=\left(
\begin{array}{c c c c}
\sigma_Q^2 & 0 & 0 & 0\\
0 & \sigma_\alpha^2 & 0 & 0\\
0 & 0 & \sigma_\sfm^2 & 0\\
0 & 0 & 0 &  \sigma^2_{\beta_a}
\end{array}
\right)=\left(
\begin{array}{c c c c}
0.01 & 0 & 0 & 0\\
0 & 100 & 0 & 0\\
0 & 0 & 0.01 & 0\\
0 & 0 & 0 & 0.01
\end{array}
\right)$$
and applies the Metropolis-Hastings rule to accept or reject them. This demands the evaluation of the likelihood $\ell({\bf y}|{\bf \theta'})$ for each candidate ${\bf \theta'}$, and therefore it is approximated using a BF, $\ell({\bf y},{\bf \theta'})\simeq \ell^N({\bf y}|{\bf \theta'})$ with $N=100$, the same as for the proposed NPMC method. We have used the algorithm to generate Markov chains of length $15\times 400=6000$, which is equivalent to the computational cost of the NPMC method with $M=400$ samples per iteration and $k=15$ iterations.

We have compared the ABC SMC algorithm, the PMH method and the proposed NPMC scheme with $k=15$ iterations and $M\in\{50,200\}$ samples in terms of the empirical normalised mean square error (NMSE) of the posterior-mean estimators of each parameter ($Q$, $\sfm$, $\alpha$ and $\beta_a$), computed from $45$ independent simulation runs.

\begin{figure*}[htb]
\centering \subfigure[]
	{\includegraphics[width=0.33\linewidth]{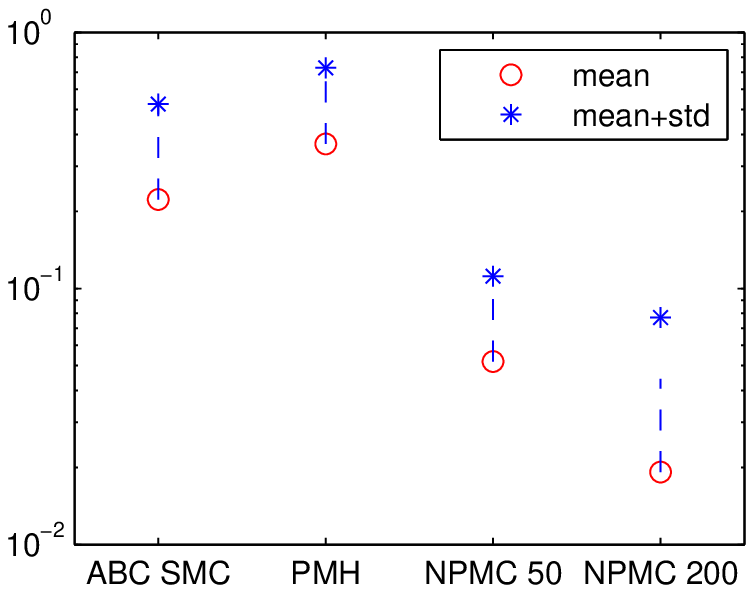}}
\subfigure[]
	{\includegraphics[width=0.33\linewidth]{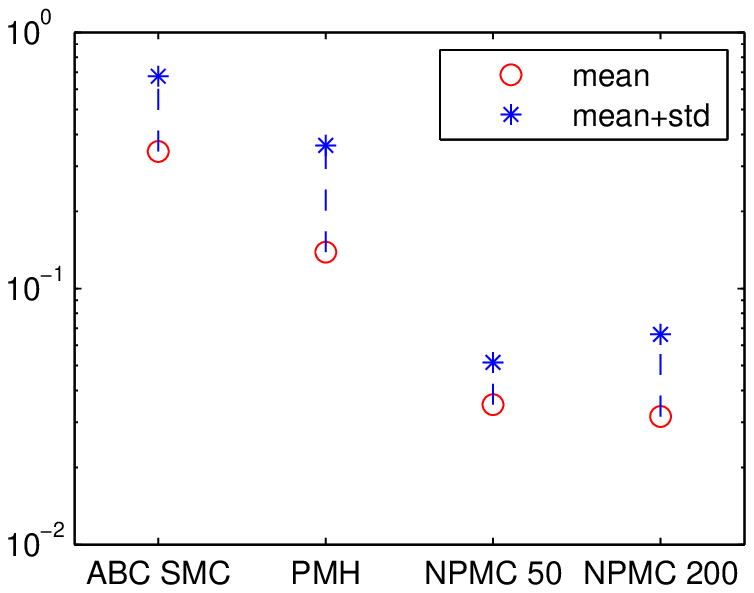}}
\subfigure[]
	{\includegraphics[width=0.33\linewidth]{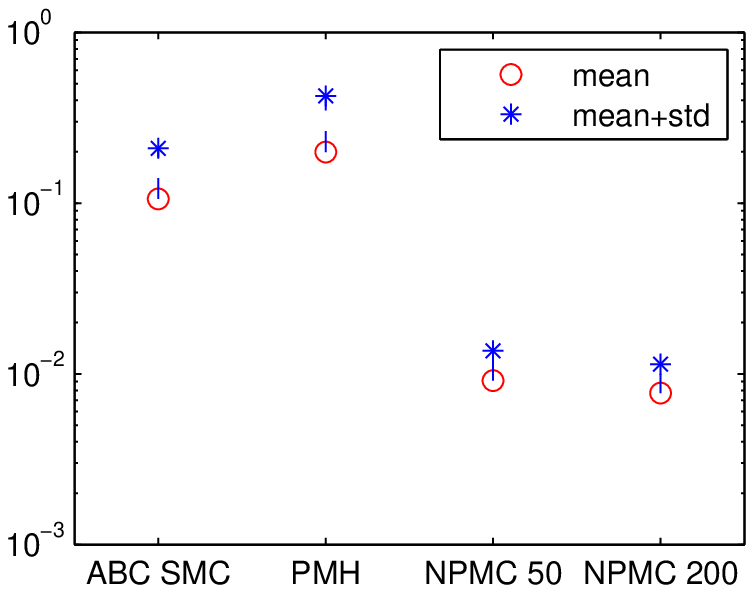}}
\subfigure[]
	{\includegraphics[width=0.33\linewidth]{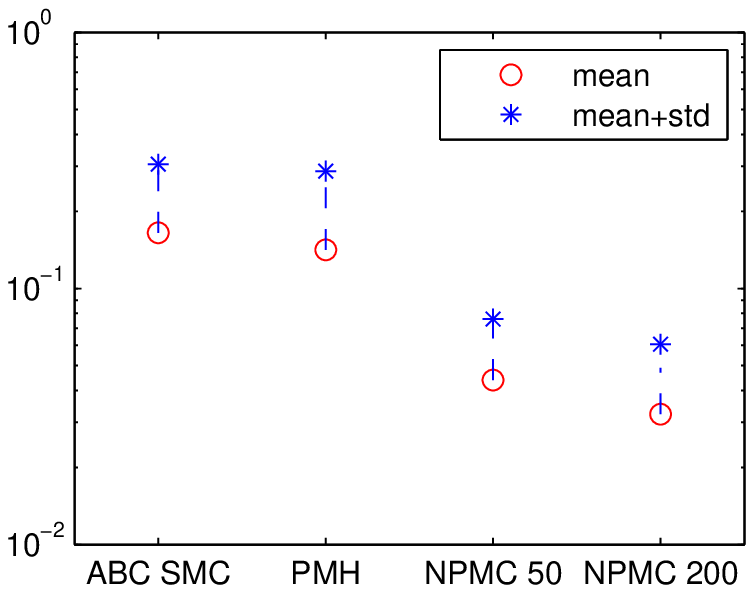}}
\caption{Average values (plus one standard deviation) of the empirical mean square errors over independent 33 simulations for all three methods (ABC SMA, PMH and NPMC with $M=50$ and $M=200$ Monte Carlo samples) and each unknown parameter:  (a) parameter $Q$, (b) parameter $\sfm$, (c) parameter $\alpha$, (d) parameter $\beta_a$.}
\label{MSE}
\end{figure*}

Figure \ref{MSE}  depicts the outcomes of the comparison. For each parameter, we shows the NMSE (and the NMSE plus one standard deviation) obtained empirically from the simulations, for each algorithm. The NPMC method, even with just $M=50$ samples, outperforms the PMH and ABC SMC techniques, even if the latter have a significantly larger computational cost. The NPMC algorithm with $M=200$ samples is still computationally less demanding than the implemented PMH and ABC SMC methods\footnote{The running time of the ABC SMC algorithm is random because it is based on a rejection sampling scheme. In the average, its computational cost is similar to an NPMC method with $400$ samples, although some simulation runs can be much faster.} and it attains an improvement in estimation accuracy. This improvement is modest for some parameters, but this is again a consequence of the limited amount of data available for the estimator.

%
\section{Conclusion} \label{sConclusions}

We have proposed a stochastic version of the coupled repressilator model of \cite{Garcia-Ojalvo04} that enables 
\begin{itemize}
\item[(a)] a mathematically principled manner of describing experimental uncertainties in the synthesis of multicellular clocks, and 
\item[(b)] the design of probabilistic methods for the estimation of unknown parameters in the model.
\end{itemize}
In particular, we have introduced an iterative Monte Carlo sampling scheme for the approximation of the posterior probability distributions of the parameters of interest conditional on a set of noisy and partial observations of the system. The technique relies on a methodology termed nonlinear importance sampling, originally introduced in \cite{Koblents14}. In this paper, we have applied this methodology to the stochastic repressilator and proved a new theoretical result regarding the convergence of nonlinear importance samplers. The new convergence theorem is stronger than the original results in \cite{Koblents14} and holds for a broader class of models (of which the stochastic repressilator system is just an instance). Specifically, we have proved that nonlinear importance samplers can converge asymptotically with optimal Monte Carlo error rates even when the importance weights can only be estimated (and the variance of these estimates is positive and cannot be reduced). 

The theoretical analysis has been complemented with an extensive computer simulation study that illustrates the relationship between the deterministic and stochastic repressilator models and demonstrates the efficiency and accuracy of the proposed estimation algorithm compared to other computational approaches of similar complexity that can be found in the literature.


\section*{Acknowledgments}

This research has been partially supported by the Spanish Ministry of Economy and Competitiveness (projects TEC2012-38883-C02-01 COMPREHENSION and FIS2013-40653-P), the Spanish Ministry of Education, Culture and Sport (mobility award PRX15/00378), the Office of Naval Research (ONR) Global (Grant Award no. N62909-15-1-2011), the regional Government of Madrid (program CASI-CAM-CM S2013/ICE-2845) and the Cancer Research UK and the Eve Appeal Gynaecological Cancer Research Fund (grant ref. A12677) supported by the National Institute for Health Research (NIHR) University College London Hospitals (UCLH) Biomedical Research Centre. 

\appendix
%
\section{The bootstrap filter} \label{apBF}

A Markov state-space model consists of two sequences of r.v.'s, $\{\bx_n\}_{n\ge 0}$ and $\{\by_n\}_{n \ge 1}$. The first sequence, $\{\bx_n\}$, is termed the system state. We assume it takes values on some space $\mX \subseteq \Real^{d_x}$, hence $\bx_n$ is a random $d_x \times 1$ vector. The state dynamics are described by a prior probability measure $\mK_0(\sd\bx_0)$ and a sequence of Markov kernels $\mK_{n,\theta}(\sd\bx_n|\bx_{n-1})$ that depend on a parameter vector $\theta \in \Real^{d_\theta}$. In the case of the modified stochastic repressilator model, the parameter vector is\footnote{We are interested on the parameters to be estimated alone. Known parameters are implicitly included in the model.} $\theta=(Q,m,\beta_a,\alpha)$ and the Markov kernel $\mK_{n,\theta}$ is given by Eq. \eqref{eqKcomposite}.

The state $\bx_n$ cannot be observed directly. Instead, some partial noisy observations $\by_n$ are collected. We assume that the observations are conditionally independent given the system states and the parameter vector $\theta$, with a conditional pdf (with respect to the Lebesgue measure) $l_{n,\theta}(\by_n|\bx_n) > 0$, which depends on the parameters $\theta$ as well. For the stochastic repressilator model, the observations are given by Eq. \eqref{eqObserv}, hence $l_{n,\theta}(\by_n|\bx_n) = l_n(\by_n|\bx_n)$ is independent of the parameter vector $\theta$ in the case of the repressilator model of interest in this paper. 

The bootstrap filter (BF) \cite{Gordon93,Doucet01} is a recursive Monte Carlo algorithm for the approximation of the sequence of posterior probability measures $\pi_{n,\theta}(\sd\bx_n)$, $n = 1, 2, ...$, where 
\begin{itemize}
\item for a {\em given} (i.e., fixed, even if arbitrary) sequence of observations $\by_{1:n}=\{\by_1, \by_2, ..., \by_n\}$,
\item and a Borel set $A \subset \mX$,
\end{itemize}
$\pi_{n,\theta}(A)$ is the probability of the even $\bx_n \in A$ conditional on the observations $\by_{1:n}$ and the parameter values given by $\theta$. The BF with $N$ {\em particles} (i.e., Monte Carlo samples) can be briefly outlined as follows.
\begin{enumerate}
\item \textit{\textbf{Initialisation.}} Draw $N$ samples $\bx_0^1, \ldots, \bx_0^N$ from the prior distribution $\mK(\sd\bx_0)$. The particle approximation of $\pi_{0,\theta}(\sd\bx_0) \equiv \mK_0(\sd\bx_0)$ is
$$
\pi_{0,\theta}^N(\sd\bx_0) = \frac{1}{N} \sum_{i=1}^N \delta_{\bx_0^i}(\sd\bx_0),
$$
where $\delta_{\bx_0^i}$ denotes the Dirac (unit) delta measure centred at $\bx_0^i \in \mX$.

\item \textbf{\textit{Recursive step.}} Given the approximation $\pi_{n-1,\theta}^N(\sd\bx_{n-1}) =  \frac{1}{N} \sum_{i=1}^N \delta_{\bx_{n-1}^i}(\sd\bx_{n-1})$, take the following steps:
	\begin{enumerate}
	\item Randomly propagate each particle using the Markov kernel in the model, i.e., draw $\tilde \bx_n^i$ from $\mK_{n,\theta}(\sd\bx_n|\bx_{n-1}^i)$, $i = 1, ..., N$.
	\item Compute IWs, $\tilde u_n^i = l_{n,\theta}(\by_n|\bx_n^i)$, for $i=1, ..., N$, and
	\item normalise them as
	$$
	u_n^i = \frac{ \tilde u_n^i }{ \sum_{j=1}^N \tilde u_n^j }, \quad i=1, ..., N.
	$$ 
	\item Resample: draw $N$ times independently from the discrete distribution $\tilde \pi_{n,\theta}^N(\sd\bx_n) = \sum_{i=1}^N u_n^i\delta_{\tilde \bx_n^i}(\sd\bx_n)$ and denote the resulting samples as $\{ \bx_n^i \}_{i=1}^N$. Construct the to {\em unweighted} approximation $\pi_{n,\theta}^N(\sd\bx_n) = \frac{1}{N} \sum_{i=1}^N \delta_{\bx_n^i}(\sd\bx_n)$.
	\end{enumerate}
\end{enumerate}

The resampling step (d) above can be implemented in a number of different ways (see, e.g., \cite{Douc05,Bain08} or \cite{Cappe07} for a brief survey of methods). Here, for simplicity, we have adopted a scheme which is often referred to as multinomial resampling \cite{Doucet00,Douc05} but most asymptotic convergence results hold true for several other schemes as well \cite{DelMoral04,Bain08}. The random measure $\pi_{n,\theta}^N$ can be used to approximate integrals w.r.t. the true posterior measure $\pi_{n,\theta}$. To be specifc, it can be shown under mild assumptions on the state space model \cite{DelMoral04,Bain08} that $\| (f,\pi_{n,\theta}^N) - (f,\pi_{n,\theta}) \|_p \le \frac{C}{\sqrt{N}}$ for every $p \ge 1$, where $f(\bx_n)$ is a real bounded function of the state $\bx_n$, $\| Z \|_p = (E[Z^p])^{\frac{1}{p}}$ is the $L_p$ norm of the r.v. $Z$, $(f,\pi) := \int f(\bx)\pi(\bx)$ denotes the integral of function $f$ w.r.t. the measure $\pi$ and $C$ is a finite constant (independent of $N$).  

The algorithm also produces an approximation of the predictive probability measure of $\bx_n$ conditional on the observations $\by_{1:n-1}$. We denote the actual predictive measure as $\xi_{n,\theta}(\sd\bx_n)$ and its $N$-particle approximation as
$$
\xi_{n,\theta}^N(\sd\bx_n) = \frac{1}{N}\sum_{i=1}^N \delta_{\tilde \bx_n^i}(\sd\bx_n).
$$ 
If we write $\by=\by_{1:n}$, it turns out that the likelihood of the parameter vector $\theta$, namely $\ell(\by|\theta)$ can actually be expressed in terms of integrals w.r.t. to the sequence of predictive distributions $\xi_{n,\theta}$. To be specific,
\begin{equation}
\ell(\by|\theta) = \prod_{k=1}^t (l_{k,\theta}(\by_k|\cdot),\xi_{k,\theta})
\label{eqDefEll}
\end{equation}
and, therefore, the bootstrap filter yields the straightforward estimator
\begin{equation}
\ell^N(\by|\theta) = \prod_{k=1}^t (l_{k,\theta}(\by_k|\cdot),\xi_{k,\theta}^N) = \frac{1}{N^t} \prod_{k=1}^t \sum_{i=1}^N l_{k,\theta}(\by_k|\tilde \bx_k^i),
\label{eqUBEstimator} 
\end{equation}
which can be proved to be unbiased under very mild assumptions \cite{DelMoral04}.

\vspace{-0.1cm}
\bibliographystyle{plain}
\bibliography{bibliografia}

\end{document}